\documentclass[preprint,11pt]{elsarticle}

\usepackage[T1]{fontenc} 
\usepackage{amssymb}
\usepackage{amsmath}
\usepackage{amsthm}
\usepackage{lineno}
\usepackage{xcolor}
\usepackage[margin=1in]{geometry}
\usepackage{pythonhighlight}
\usepackage{algorithm}
\usepackage[noend]{algpseudocode}
\usepackage{tabularx}
\usepackage[utf8]{inputenc}
\usepackage[T1]{fontenc}
\usepackage{textcomp}
\usepackage{underscore}
\usepackage{mathtools}
\usepackage{hyperref}
\usepackage{booktabs}

\usepackage{nomencl}
\makenomenclature

\theoremstyle{plain}

\theoremstyle{definition}

\theoremstyle{remark}

\usepackage{lineno}

\journal{Computational Materials Science}

\begin{document}

\begin{frontmatter}

\title{PolyGraphPy: A unified Python framework for atomistic simulation and machine learning-driven polymer design}

\author[label1]{João G. C. S. Duarte}
\author[label3]{Shruti Venkatram}
\author[label3]{Morgan Cencer}
\author[label1,label2]{Traian Dumitricǎ}
\author[label1]{Ketson R. M. dos Santos}

\address[label1]{Department of Civil, Environmental, and Geo- Engineering, University of Minnesota, Minneapolis, MN, 55455-0116, USA}
\address[label2]{Department of Mechanical Engineering, University of Minnesota, Minneapolis, MN, 55455-0116, USA}
\address[label3]{3M Company, 3M Center, St. Paul, MN, 55144-1000, USA}

\begin{abstract}
  Polymers are indispensable materials with daily applications ranging from electronics to medicine, owing to their versatility, which can be tailored by adjusting their chemical composition and architecture. The design space for these compounds is vast and governed by factors such as monomer classes, copolymer configurations (e.g., linear, branched, random, and alternating), chain size, stoichiometry, and material properties (e.g., density, refractive index, solubility, and Poisson’s ratio). Thus, its exploration requires efficient computational methodologies for polymer science. To address this challenge, this paper introduces PolyGraphPy, an open-source, unified Python framework that integrates atomistic simulations with machine learning for accurate property prediction and property-guided polymer design. The framework automates quantum mechanics calculations using Density Functional Tight Binding (DFTB+) to efficiently construct structured datasets for monomers, homopolymers, and alternating copolymers. For property prediction, PolyGraphPy employs Bayesian Graph Neural Networks (GNNs) utilizing stochastic graph representations to predict target properties—such as static polarizability, while providing robust uncertainty quantification. Furthermore, the platform incorporates two complementary generative models for the \textit{de novo} design of targeted molecules: a SELFIES-based Generative Pretrained Transformer (GPT) and a Genetic Algorithm (GA) based on BRICS graph fragmentation. Demonstrated on a dataset of acrylates, PolyGraphPy provides a highly customizable, end-to-end pipeline that reduces computational costs and accelerates data-driven polymer informatics.
\end{abstract}


\end{frontmatter}

\section{Introduction}
Recently, machine learning (ML) has emerged as a potent tool for enabling the discovery of new polymers due to its ability to learn chemical patterns from data. However, identifying an appropriate polymer representation remains a central challenge in developing ML models for polymer property prediction and property-guided generation \cite{Wu2020, Peerless2019}. To this end, several text-based representations tailored to polymer chemistry have been proposed, including SMILES \cite{Weininger1988}, BigSMILES \cite{Lin2019}, InChI \cite{Heller2015}, and PolyGrammar \cite{Guo2022}. Nevertheless, learning their semantics is challenging for ML models, and generative models often produce invalid molecules due to the fragile syntax of these representations, where minor structural errors lead to chemically invalid strings. Alternatively, SELFIES (Self-Referencing Embedded Strings) \cite{Krenn_2020} provides a robust string representation of molecules in a surjective manner, where each SELFIES string corresponds to a valid molecule. Furthermore, a graph representation offers a natural manner to encode molecular structures, where atoms and bonds are modeled as nodes and edges, respectively \cite{Aldeghi2022}. 

Polymer informatics is a highly interdisciplinary field that integrates ML and atomistic simulations to accelerate both property prediction and property-guided design of polymers. Several computational tools have been developed to enable efficient analysis and design of polymers. Among the existing ones, the Polymer Genome \cite{Kim2018, Tran2020} is a web-based ML platform for near-instantaneous prediction of polymer physicochemical properties. Built with Python and PHP, it provides an intuitive interface and a robust backend infrastructure for rapid screening of candidate polymers; however, it remains strictly database-bound, relying on pre-existing historical datasets rather than executing dynamic, automated quantum simulations to explore unmapped chemical spaces. Additionally, Chemprop is a versatile ML package that employs a Directed Message Passing Neural Network (D-MPNN) architecture for property prediction. Initially developed for small molecules, Chemprop has since been extended to polymers \cite{Yang2019, Aldeghi2022, Heid2024, Ignacz2025, Sun2025}. While highly precise for property estimation, Chemprop functions strictly as a predictive framework and lacks an integrated loop for automated \textit{de novo} generation or structural optimization. DeepMol is an open-source framework that leverages automated machine learning (AutoML) to predict molecular physical and chemical properties \cite{Correia2024}. espite its flexibility across varied pipelines, DeepMol is restricted to standard classification and regression architectures without built-in molecular search heuristics or generative agents. Polymerexpert \cite{Bicerano2024} is a system designed to support \textit{de novo} polymer design by integrating domain knowledge (polymer chemistry heuristics, synthesizability rules) with data-driven predictions of polymer properties. Furthermore, XenonPy \cite{Yamada2019} serves as a comprehensive materials informatics library that excels in computing polymer-specific compositional and structural descriptors for ML mapping, though it lacks dynamic structural generation. Similarly, PolyGNN \cite{Gurnani2023} provides an advanced graph-based predictive architecture tailored specifically for polymer networks, but it operates without integrated active-learning uncertainty quantification or \textit{de novo} design agents.

The rapid development of generative tools has further accelerated the discovery of new chemical species and materials. In this regard, the package SPACIER \cite{Nanjo2025} is a fully automated system that integrates all-atom classical MD simulations with ML to enable property-guided molecular generation via Bayesian optimization. While SPACIER successfully couples simulation with active exploration, it operates exclusively in the classical molecular dynamics (MD) regime, omitting electronic-level quantum mechanics optimizations, and relies on structural screening rather than a dual-engine structural generation approach. In the realm of generative design, platforms like REINVENT \cite{Blaschke2020} have set a benchmark for molecular generation using Recurrent Neural Networks (RNNs) and reinforcement learning. However, while highly effective for small drug-like molecules, they do not natively orchestrate automated quantum-level simulations or target macroscopic polymer attributes. SELF-BART \cite{Priyadarsini2024} is an encoder–decoder model built upon the BART architecture \cite{lewis2019}, designed to learn robust molecular representations and generate novel molecules directly from SELFIES strings. This approach combines the syntactic robustness of SELFIES with the generative power of BART to effectively explore valid chemical space. In addition, genetic-guided GFlowNets \cite{kim2024} combine generative flow networks with domain-informed genetic algorithms, allowing the generative policy to incorporate expert knowledge and improve sample efficiency. MG$^2$N$^2$ \cite{Bongini_2021} adopts a sequential graph generation strategy based on graph neural networks (GNNs), incrementally constructing molecules by adding nodes (atoms or substructures) and corresponding edges, thus enabling a flexible and interpretable molecular generation process. These examples illustrate the expanding ecosystem of data-driven platforms that are transforming polymer research by significantly reducing the time and computational cost of materials discovery and optimization.

To highlight the unique value proposition of the proposed work amidst this expanding landscape, Table~\ref{tbl:benchmarking} systematically benchmarks \texttt{PolyGraphPy} against existing prominent frameworks. This comparison demonstrates that \texttt{PolyGraphPy} represents a distinct methodological consolidation that addresses key operational gaps in automated quantum simulation, uncertainty quantification, and structural discovery.

\begin{table}[htbp]
  \small
  \caption{Comparative benchmarking of \texttt{PolyGraphPy} against existing software frameworks in polymer informatics.}
  \label{tbl:benchmarking}
  \resizebox{\textwidth}{!}{
  \begin{tabular}{lccccc}
    \hline
    \textbf{Framework} & \textbf{Quantum} & \textbf{Predictive} & \textbf{Uncertainty} & \textbf{Generative} & \textbf{Target Application} \\
    & \textbf{Simulation} & \textbf{Architecture} & \textbf{Quantification} & \textbf{Capabilities} & \textbf{Space} \\
    \hline
    Polymer Genome \cite{Kim2018} & No (Static DB) & ML Regression / NN & No & No & General Polymers \\
    Chemprop \cite{Yang2019} & No & D-MPNN & Optional (Ensemble) & No & Molecules / Polymers \\
    DeepMol \cite{Correia2024} & No & AutoML / Descriptors & No & No & Bio/Chemical Systems \\
    XenonPy \cite{Yamada2019} & No & NN / Transfer Learning & No & No & Materials / Polymers \\
    PolyGNN \cite{Gurnani2023} & No & Graph Neural Network & No & No & Polymers \\
    SPACIER \cite{Nanjo2025} & No (Classical MD) & Gaussian Process & Yes & Screening Only & General Materials \\
    REINVENT \cite{Blaschke2020} & No & RNN / RL Proxy & No & Yes (RNN/RL) & Small Molecules \\
    \hline
    \textbf{\texttt{PolyGraphPy}} & \textbf{Yes} & \textbf{Bayesian} & \textbf{Yes} & \textbf{Dual Engine} & \textbf{Acrylates, Homo-,} \\
    \textbf{(This Work)} & \textbf{(Automated DFTB+)} & \textbf{GraphUNet} & \textbf{(MC Dropout)} & \textbf{(GPT \& BRICS-GA)} & \textbf{\& Copolymers} \\
    \hline
  \end{tabular}
  }
\end{table}

Given the current advances in ML-based polymer analysis and design, we propose an open-source Python-based computational framework, \texttt{PolyGraphPy}, for efficient and accurate property prediction of polymers, combined with property-guided generation. The proposed Python toolbox leverages density functional tight binding (DFTB) quantum mechanics calculations \cite{Elstner2007, dftb2020} to efficiently build a large dataset for training graph-based predictive models endowed with uncertainty quantification capabilities and two generative models: one SELFIES-based generative pretrained transformer (GPT) model \cite{radford2019language} and (ii) a property-guided genetic algorithm (GA) \cite{Holland1992, Katoch2021}. The \texttt{PolyGraphPy} toolbox, including all scripts and examples discussed in this manuscript, is freely available on GitHub at \href{https://github.com/SASP-lab/PolyGraphPy}{https://github.com/SASP-lab/PolyGraphPy}.

\section{Materials and Methods}
\subsection{Computational Framework Overview}
\texttt{PolyGraphPy} is a Python toolbox for the efficient property prediction and property-guided design of polymers, which integrates the following tools into a unified platform:
\begin{enumerate}
    \item DFTB-based quantum mechanics calculations through a user-friendly interface with efficient atomistic simulations that leverage parallel processing to enhance its computational performance. 
    \item Homopolymer and copolymer property prediction and uncertainty quantification with Bayesian graph neural networks \cite{Scarselli2009, kipf2016, gao2019}.
    \item Property-guided generation of homopolymers using a SELFIES-based GPT \cite{Krenn_2020} and a GA based on graph fragmentation using breaking of retro-synthetically interesting chemical substructures (BRICS) decomposition algorithm \cite{Degen2008}.
\end{enumerate}

\begin{figure}
    \centering
    \includegraphics[width=\linewidth]{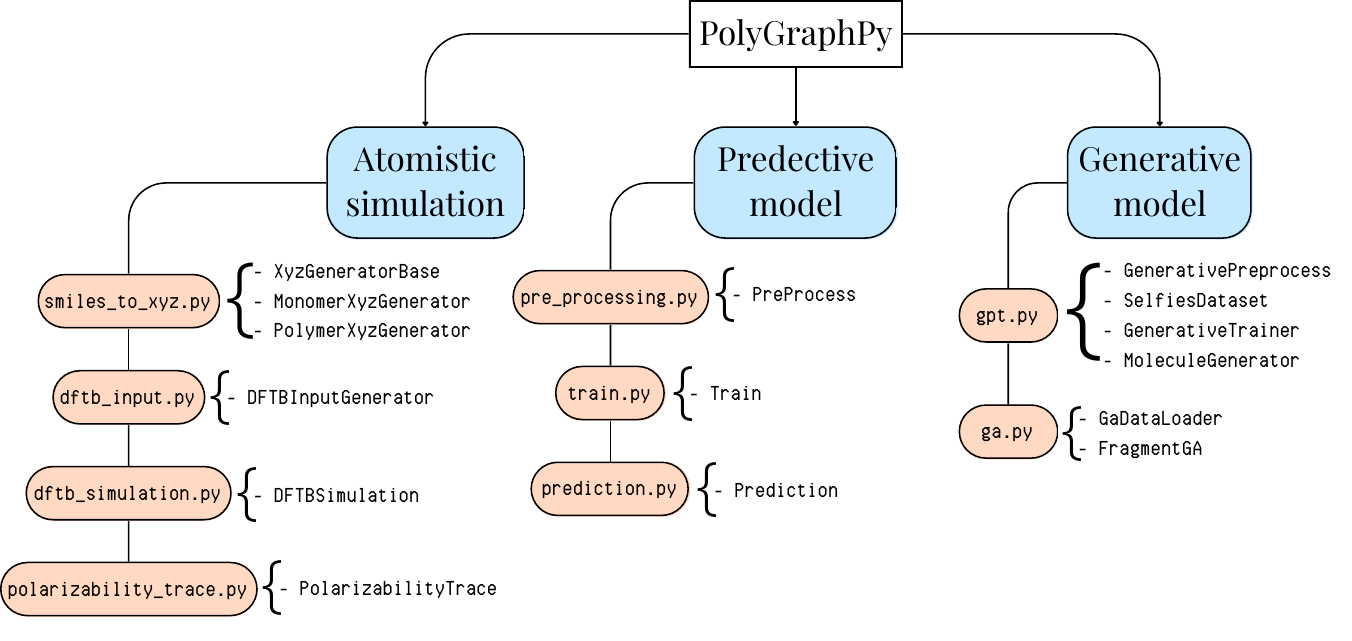} \caption{Architecture, core Python scripts, and associated classes of \texttt{PolyGraphPy}.}
    \label{fig:folders_and_classes}
\end{figure}

\texttt{PolyGraphPy} is based on PyTorch \cite{paszke2019pytorchimperativestylehighperformance} and PyTorch Geometric \cite{fey2019fastgraphrepresentationlearning}, and its implementation follows an object-oriented programming (OOP) paradigm, allowing efficient maintenance and rapid adaptation and retraining for the analysis of polymeric materials and small molecules. The main modules of \texttt{PolyGraphPy} (i.e., atomistic simulation, predictive model, and generative model) and their classes are shown in Fig. \ref{fig:folders_and_classes}.  

The atomistic simulation module provides user-friendly encapsulation for quantum mechanics computations employing the third-party DFTB+ software \cite{DFTBplus, Hourahine2025}. To this end, a set of smiles is provided and converted into the corresponding \texttt{.xyz} file with \texttt{smiles\_to\_xyz.py}, which serves to build the input file of DFTB+ with \texttt{dftb\_input.py}. In addition, the DFTB start is initiated with \texttt{dftb\_simulation.py}. Although any physical-chemical property can be considered in the analysis and design of polymers with \texttt{PolyGraphPy}, this paper focuses on the determination of the static polarizability of acrylates. To this end, \texttt{polarizability\_trace.py} extracts the necessary information from the results of the atomistic simulations performed with DFTB+ to calculate the static polarizability. 

The module containing the predictive model is composed of classes implemented in \texttt{pre\_processing.py}, \texttt{train.py}, and \texttt{prediction.py}, which are responsible for the operationalization of the proposed Bayesian GNN model. Furthermore, two proposed generative models, the SELFIES-based GPT (\texttt{gpt.py}) and the property-guided GA (\texttt{ga.py}), are included in a dedicated module. It is also important to emphasize that the predictive model is integrated into the generative module to determine the acceptance or rejection of generated molecules.

The modularity of this open-source platform makes it highly customizable, allowing users to retrain the developed models to meet their specific objectives. For example, hybrid generative models can be easily constructed by combining existing methods. In addition, active learning methodologies can be developed and integrated into local implementations of the proposed computational framework. 

\subsection{Atomistic Simulations}
The accuracy offered by Density Functional Theory (DFT) \cite{Hohenberg1964, Kohn1965} makes it an attractive choice for reliable quantum mechanics simulations of molecules. However, its inherent high computational cost limits its application in the construction of structured datasets from atomistic simulations of polymers. In contrast, force field (FF) methods \cite{Patil2020} provide a more efficient approach for polymer simulations but lack an explicit representation of the electronic structure, which can result in inaccurate estimations of polymer properties. Alternatively, the semi-empirical DFTB method \cite{dftb2020} approximates the electronic structure of molecules using precomputed DFT Slater–Koster (SK) parameters, offering a substantial improvement in computational efficiency while retaining quantum-mechanical fidelity. This capability makes DFTB a promising tool for efficiently generating large, well-structured datasets for training machine learning models. Consequently, integrating DFTB+ into \texttt{PolyGraphPy} provides seamless access to a powerful platform for efficient atomistic simulations. Next, an overview of this technique is provided based on the detailed discussion in Ref.~\citenum{DFTBplus}, and it covers the main concepts related to diverse DFTB formulations and their relationship with the expansion of the total energy functional.

\subsubsection{Density Functional Tight Binding}
In the DFTB formulation, the total energy functional ($E[\rho_0]$) is expanded around a reference electron density $\rho_0$ (e.g., a superposition of neutral atomic densities), where the ground state density $\rho(\mathbf{r}) = \rho_0(\mathbf{r}) + \delta\rho(\mathbf{r})$ is determined as a perturbation of $\rho_0$, which leads to a Taylor expansion of the total energy functional in terms of the density fluctuation $\delta\rho$ \cite{DFTBplus}:
\begin{equation}\label{eq:dftb_expansion}
    E[\rho_0+\delta\rho] = E^0[\rho_0] + E^1[\rho_0,\delta\rho] + E^2[\rho_0,(\delta\rho)^2] + E^3[\rho_0,(\delta\rho)^3] + \cdots
\end{equation}
The appropriate truncation of this expansion determines different DFTB methods, such as DFTB1, DFTB2, and DFTB3 \cite{elstner2000, Elstner2007, Gaus2014, dftb2020}. The DFTB1 method is obtained by keeping only the first two terms of this expansion (e.g., $E^0[\rho_0]$, $E^1[\rho_0,\delta\rho]$), which are approximated by projecting the Kohn-Sham orbitals $\boldsymbol{\Psi}_i(\boldsymbol{r})$ onto a basis of confined atomic orbitals $\phi_{\mu}(\boldsymbol{r})$ computed with the atomic Kohn-Sham equations, while considering an additional compression with respect to the free atomic wave function. Therefore, $\boldsymbol{\Psi}_i(\boldsymbol{r})$ is written as follows: 
\begin{equation}\label{eq:KSorbitals}
    \boldsymbol{\Psi}_i(\boldsymbol{r}) = \sum_{\mu}c_{i\mu}\phi_{\mu}(\boldsymbol{r})
\end{equation}
where $c_{i\mu}$ are unknown coefficients. Considering the calculated Hamiltonian $H^0_{\mu\nu} = \langle\phi_{\mu}|\hat{H}[\rho_0]|\phi_{\nu}\rangle$, and assuming that $E^0[\rho_0]$ is determined by fast-decaying two-body repulsive potentials ($V^{rep}_{AB}$), $E^{DFTB1}=E^0[\rho_0] + E^1[\rho_0,\delta\rho]$ is obtained as follows:
\begin{equation}\label{eq:dftb1}
    E^{DFTB1} = \sum_i \sum_{AB} \sum_{\mu \in A, \nu \in B} c^{\star}_{i\mu}c_{i\nu}H^0_{\mu \nu} + \frac{1}{2}\sum_{AB}V^{rep}_{AB}
\end{equation}
In this regard, the construction of precomputed integral tables via Slater-Koster transformation \cite{Slater1954} for $\langle\phi_{\mu}|\hat{H}[\rho_0]|\phi_{\nu}\rangle$ enables an efficient determination of the Hamiltonian $H^0_{\mu\nu}$ \cite{Hourahine2025}.

The previously shown DFTB1 formulation is non-self-consistent because it depends only on a single diagonalization of the Hamiltonian matrix to determine the unknown coefficients $c_{i\nu}$ and $E^{DFTB1}$. However, including higher-order terms in the expansion leads to the construction of a self-consistent method. Therefore, keeping up until the third-order term leads to the following approximation:
\begin{equation}\label{eq:dftb3}
    E^{DFTB3} = E^{DFTB1} + \frac{1}{2}\sum_{AB}\Delta q_A \Delta q_B\gamma_{AB} + \frac{1}{3}\sum_{AB}\Delta q^2_A \Delta q_B\Gamma_{AB}
\end{equation}
where integrals are analytically estimated as two-body functions $\gamma_{AB}$ describing the interaction of atomic point charges $\Delta q_A$, where $\Gamma_{AB}$ is the derivative of $\gamma_{AB}$ with respect to atomic charges \cite{Hourahine2025}. The software DFTB+ allows users to utilize these DFTB formulations, and \texttt{PolyGraphPy} considers DFTB3 its default option. 

\subsubsection{Input data preparation and DFTB simulation}\label{input_data_geo_generation}
\texttt{PolyGraphPy} automates DFTB+ input generation using the classes \texttt{MonomerXyzGenerator} and \texttt{PolymerXyzGenerator}, which take monomer SMILES strings and generate \texttt{.xyz} files for both monomers and polymers via the \texttt{generate} method.

\begin{figure}[!ht]
    \centering
    \includegraphics[width=\linewidth]{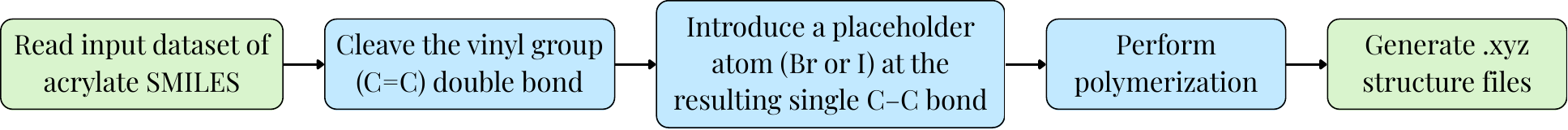}
    \caption{Steps for the construction of acrylate homopolymers.}
    \label{fig:homopolymer_process}
\end{figure}

With a focus on acrylates, two methods are implemented for the construction of the \texttt{.xyz} files of homopolymers and copolymers. For a homopolymer acrylate with a prescribed chain length (e.g., 1, 2, and 3), the construction of the \texttt{.xyz} files, as shown in Fig.~\ref{fig:homopolymer_process}, consists of cleaving the vinyl group of each monomer, enabling the introduction of a placeholder atom (e.g., Br or I) at the resulting single bond (\texttt{C--C}).

\begin{figure}[!ht]
    \centering
    \includegraphics[width=\linewidth]{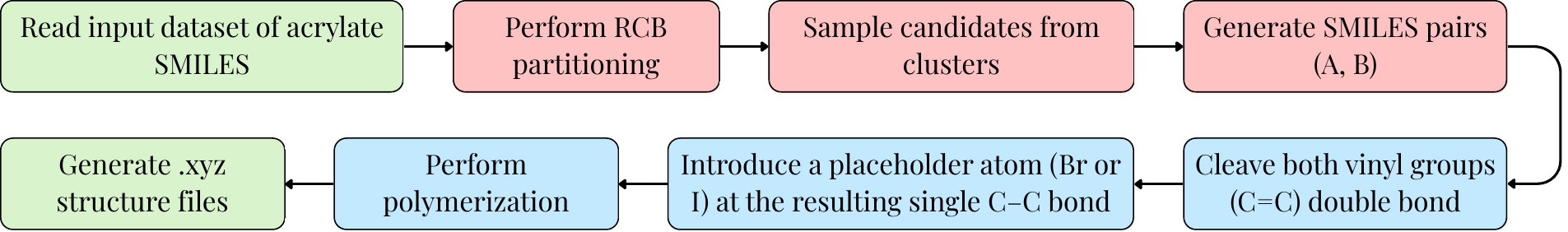}
    \caption{Steps for the construction of acrylate copolymers.}
    \label{fig:copolymer_process}
\end{figure}

In the case of alternating copolymers, the construction of \texttt{.xyz} files, illustrated in Fig.~\ref{fig:copolymer_process}, involves the pairwise combination of monomers in the form ($A,B$), where $A$ and $B$ are the selected monomers. To promote structural diversity in the constructed dataset, the monomers are first partitioned into multiple clusters using a recursive coordinate bisection (RCB) method applied to the molecular descriptors, including molecular weight, complexity, and polar surface area.

Four recursive levels are considered when running the RCB method, yielding sixteen clusters. Monomers are then randomly selected from these clusters to generate unique copolymer pairs. For each monomer in a pair ($A, B$), the vinyl groups (\texttt{C=C}) are cleaved, and placeholder halogen atoms (Br or I) are introduced at the resulting single \texttt{C--C} bonds, resembling the same process for homopolymers construction. Alternating \texttt{AB} polymer chains are subsequently assembled using the STK library\cite{turcani2018stk}, followed by geometry optimization via Hamiltonian Monte Carlo. The finalized copolymer structures are then exported as \texttt{.xyz} files.

For example, constructing a set of \texttt{.xyz} files from a collection of SMILES in \texttt{input_csv} consists of using the following commands to instantiate the classes, either for creating monomers:
\begin{python}
    xyz_creator = MonomerXyzGenerator(input_csv)
\end{python}
or polymers:
\begin{python}
    xyz_creator = PolymerXyzGenerator(input_csv, polymer_chain_size=polymer_chain_size, polymer_type=polymer_type)
\end{python}
The obtained \texttt{.xyz} files are created with the command \texttt{xyz_data = xyz_creator.generate()}, where the \texttt{polymer_chain_size} determines the number of pairs of repeating units in the chain and \texttt{polymer_type} determines the construction of monomer, homopolymer, and copolymer. In the proposed framework, each polymer size is associated with a flag $f$, which maps directly to the index in the chain list representation. The chain size $S_c$ is defined by the relation $S_c=2^f$, resulting in a size mapping of $f=0\rightarrow S_c=1$, $f=1\rightarrow S_c=2$, $f=2\rightarrow S_c=4$, and so on.

The created \texttt{.xyz} files are used to construct the input file for DFTB+ (\texttt{.hsd}) using the class \texttt{DFTBInputGenerator} and the method \texttt{prepare_input(xyz_file)}. After creating the input files, the atomistic simulations are performed by running the method \texttt{run} of the class \texttt{DFTBSimulation}, as shown in the following code snippet. 
\begin{python}
dftb_simulation = DFTBSimulation(dftbplus_path=dftbplus_path)
dftb_simulation.run()
\end{python}

\subsubsection{Postprocessing}\label{data_postprocessing}
In this paper, we use DFTB+ to calculate the electric polarizability via a coupled-perturbed linear response approach, which is expressed as follows\cite{Oviedo2010}:
\begin{equation}
    \boldsymbol{\mu}(\omega) = \boldsymbol{\alpha}(\omega)\mathbf{E}(\omega)
\end{equation}
where $\boldsymbol{\mu}(\omega)$ is the molecule's dipole moment tensor, $\boldsymbol{\alpha}(\omega)$ is the polarizability tensor, and $\mathbf{E}(\omega)$ is the static electric field. In the proposed framework, the static polarizability is assumed to be the isotropic value, and we determine the static polarizability $\alpha$ for $\omega=0$ as:
\begin{equation}
    \alpha = \frac{\alpha_{xx} + \alpha_{yy} + \alpha_{zz}}{3}
\end{equation}
where $\alpha_{xx}$, $\alpha_{yy}$, and $\alpha_{zz}$ are the entries of the diagonal of the polarizability tensor. To this end, the class \texttt{PolarizabilityTrace} can be used to estimate $\alpha$ as:
\begin{python}
static_polarizability = PolarizabilityTrace().run(input_csv)
\end{python}
It is important to mention that the proposed open-source toolbox is highly customizable and allows user-defined methods focused on diverse properties.

\subsubsection{Dataset construction}\label{data_construction}

In the \texttt{PolyGraphPy} toolbox, DFTB is used to generate a dataset of atomistic simulations for training and testing the machine learning models. The pre-processing workflow is initiated by instantiating the \texttt{PreProcess} class and executing its \texttt{run} method, as shown below:
\begin{python} 
pre_process_engine = PreProcess(input_csv=input_csv,train_input_data_path=train_input_data_path, polymer_type=polymer_type, target=prediction_target, gnn_output_path=gnn_output_path) 
data = pre_process_engine.run() 
\end{python}

The pre-processing stage begins with reading the molecular dataset, provided as a table containing the following fields: monomer identifiers (\texttt{id_A}, \texttt{id_B}); SMILES strings for each monomer (\texttt{smiles_A}, \texttt{smiles_B}); the polymer chain length (\texttt{chain_size}); the \texttt{xx}, \texttt{yy}, and \texttt{zz} components of the polarizability tensor; the static polarizability (\texttt{static_polarizability}); and the molecular type (\texttt{type}, i.e., monomer, homopolymer, or copolymer). A subsequent outlier-detection step is applied to identify anomalous entries.

Furthermore, RDKit is employed to extract atom- and bond-level features. Atom features include the atomic symbol, atomic number, degree, atomic mass, total degree, total valence, aromaticity, formal charge, and chain size. Bond features include bond type, conjugation, and aromaticity. All categorical and numerical features are encoded using a OneHotEncoder to ensure compatibility with machine-learning pipelines. The molecular structure is then converted into a graph representation using PyTorch Geometric, with atoms as nodes and bonds as edges, yielding (i) an atom-feature matrix, (ii) an edge-index matrix defining graph connectivity, (iii) matrices of bond attributes, and (iv) edge weights.

Finally, the processed molecular graphs are stored as \texttt{.pt} files for downstream use in the GNN model. Each output file contains the atom-feature matrix (\texttt{x}), edge connectivity (\texttt{edge_index}), bond attributes (\texttt{edge_attr}), target property (\texttt{y}), edge weights (\texttt{edge_weight}), molecular identifiers (\texttt{id_A}, \texttt{id_B}), and chain length (\texttt{chain_size}). For homopolymers and monomers, \texttt{id_B} is assigned \texttt{None}, while for monomers, \texttt{chain_size} is set to 0. These conventions ensure a consistent and faithful graph-based representation across all molecular types.


\subsection{GNN-Based Property Prediction}

\begin{figure}[!ht]
    \centering
    \includegraphics[width=0.7\linewidth]{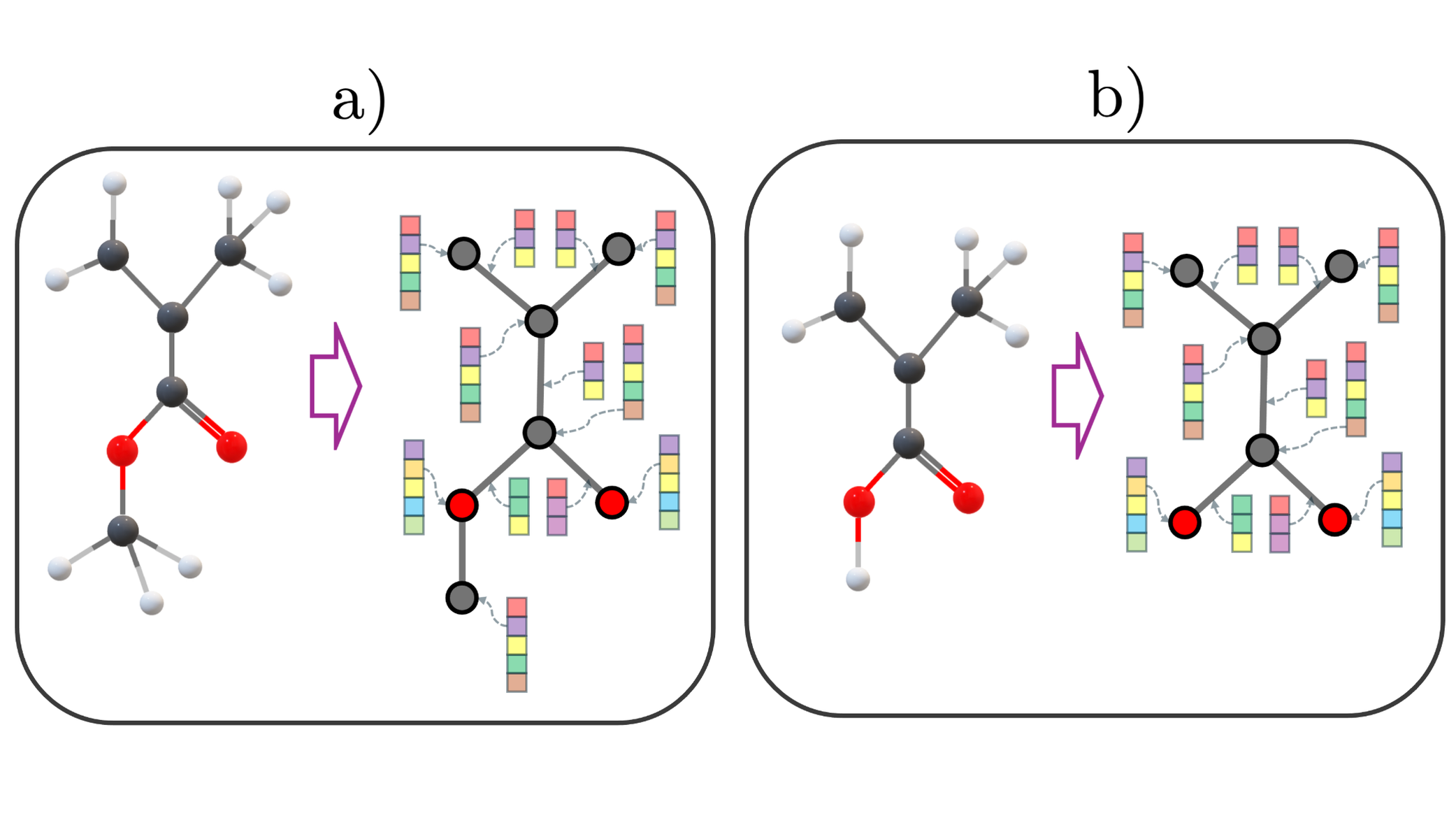}
    \caption{Graph representation of a methyl acrylate molecule. Gray circles represent hydrogen (H) atoms, black circles represent carbon (C) atoms, and red circles represent oxygen (O) atoms. Black solid lines indicate single bonds, while red solid lines denote double bonds.}
    \label{fig:graph_representation}
\end{figure}

A natural way to represent molecules is as graphs, where atoms correspond to vertices and chemical bonds to edges (see Fig.~\ref{fig:graph_representation}). This representation enables the use of methodologies developed for non-Euclidean data structures, such as graph neural networks (GNNs) \cite{Kearnes2016, Stokes2020, Jiang2021}. GNNs have achieved remarkable success in predicting a wide range of polymer properties \cite{zeng2018, Gurnani2023, wang2024}, providing a powerful framework for learning directly from molecular graph structures.

\begin{figure}
    \centering
    \includegraphics[width=0.75\linewidth]{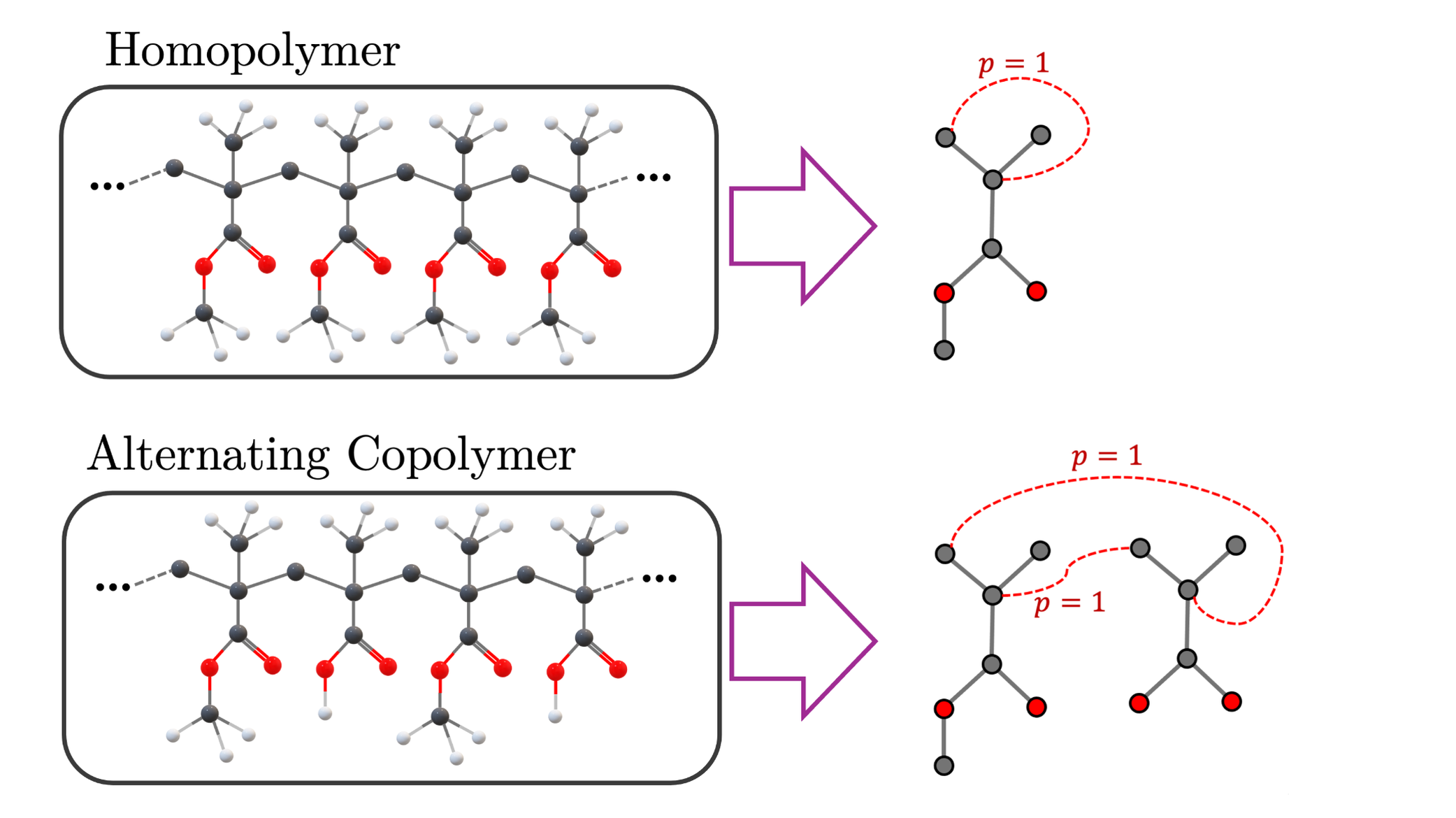}
    \caption{Stochastic connection representation showing the frequency of links between repeating units (e.g., AA for homopolymers, AB for copolymers). Stochastic edges represent bonds weighted by their occurrence frequency in the polymer chain: $w = 1.0$ for homopolymers and $w = 0.5$ for alternating copolymers.}
    \label{fig:polymers_weight}
\end{figure}

In this work, we adopt a graph-based representation of polymers following the strategy proposed by \cite{Aldeghi2022}, which enables an efficient encoding of non-structured data structures representing polymers and accurate prediction of properties of interest. This representation is particularly suitable for polymers, as it naturally captures the sequence of repeating units through the use of stochastic bonds corresponding to edges that mathematically describe the frequency with which the repeating units appear in the polymer chain. Each stochastic bond is assigned a weight $w\in(0,1]$, which can connect either identical monomers (e.g., A–A) to represent homopolymers or distinct monomers (e.g., A–B) to represent copolymers, as illustrated in Fig.~\ref{fig:polymers_weight}. Therefore, uncovering hidden patterns within a complex, unstructured dataset of graphs requires the use of appropriate computational tools. To this end, we propose a Bayesian graph convolutional network (GCN) (see \cite{kipf2017} for a discussion on graph convolutions) to enable the accurate prediction of the properties of polymers.

\subsubsection{Bayesian graph neural networks}

Any neural network is a model designed to be accurate and reliable. However, the limited knowledge about the functional relationship between the molecular structure and the property of interest (i.e., static polarizability) undermines the reliability of predictions, which is determined through the appropriate quantification of epistemic uncertainties \cite{Wasserman2000}. Therefore, determining how this uncertainty influences the predictions made with the trained GNN models used in this work is a necessary step for ensuring the reliability of the obtained outcomes. 

Mathematically, a GNN is represented as a model $\boldsymbol{\alpha} = f(\mathbf{G};\mathbf{w})$---where $\mathbf{w}$ are the GNN weights---that maps the input graphs $\mathbf{G} = \{\mathcal{G}_1, \mathcal{G}_2, \dots, \mathcal{G}_N\}$ to output properties $\boldsymbol{\alpha} = \{\alpha_1, \alpha_2, \dots, \alpha_N\}$; which, without loss of generality, can be assumed to be the static polarizability $\alpha$. Therefore, we aim to determine the predictive probability $p(\alpha^{*}|\mathcal{G}^{*},\mathbf{G},\boldsymbol{\alpha})$ of $\alpha^{*}$ for an unseen input graph $\mathcal{G}^{*}$, given the training dataset $\{\mathbf{G},\boldsymbol{\alpha}\}$. This probability is approximated as follows:
\begin{equation}\label{eq:prob1}
    p(\alpha^{*}|\mathcal{G}^{*},\mathbf{G},\boldsymbol{\alpha}) \approx \int p(\alpha^{*}|f)\,p(f|\mathcal{G}^{*},\mathbf{w})p(\mathbf{w}|\mathbf{G},\boldsymbol{\alpha})\,dfd\mathbf{w}.
\end{equation}

As the direct calculation of the posterior distribution $p(\mathbf{w}|\mathbf{G},\boldsymbol{\alpha})$ is intractable due to the dimensionality of the integral, a variational approach is sought as an alternative. To this end, we approximate this posterior distribution with a variational distribution $q(\mathbf{w})$ that minimizes the Kullback–Leibler (KL) divergence $\mathrm{KL}\!\left[q(\mathbf{w}) \,\|\, p(\mathbf{w}|\mathbf{G},\boldsymbol{\alpha})\right]$~\cite{Blei2017}. The predictive probability then becomes
\begin{equation}\label{eq:prob3}
    p(\alpha^{*}|\mathcal{G}^{*}) \approx \int p(\alpha^{*}|f)\,p(f|\mathcal{G}^{*},\mathbf{w})\,q(\mathbf{w})\,df\,d\mathbf{w}.
\end{equation}

Assuming the model form $f(\cdot,\mathbf{w})$ is known (i.e., either GNN or MLP with weights $\mathbf{w}$), the propagation of uncertainty from $\mathbf{w}$ to the static polarizability $\alpha^{*}$ is determined as follows:
\begin{equation}\label{eq:prob4}
    p(\alpha^{*}|\mathcal{G}^{*}) \approx \int p(\alpha^{*}|\mathcal{G}^{*},\mathbf{w})\,q(\mathbf{w})\,d\mathbf{w}.
\end{equation}
This equation forms the theoretical foundation for Bayesian neural networks. Therefore, to determine the uncertainty in the predicted static polarizability for a given graph $p(\alpha^{*}|\mathcal{G}^{*})$, we assume that the GNN $\boldsymbol{f}(\cdot,\mathbf{w})$ is a neural network with random weights.

Since the true variational distribution $q(\mathbf{w})$ is generally unknown, it is often approximated using a regularization scheme applied to the network’s point estimates, effectively treating the regularization as a prior for $q(\mathbf{w})$~\cite{yarin2015,jospin2022}. A common variational approximation assumes that the weights $\mathbf{w}$ follow a Bernoulli distribution:
\begin{equation}\label{eq:bernoulli}
    \mathbf{z}_{i,j} \sim \mathrm{Bernoulli}(p_j) =
    \begin{dcases}
        1, & \text{with probability } p_j,\\[4pt]
        0, & \text{with probability } 1 - p_j,
    \end{dcases}
\end{equation}
where $j = 1, \dots, n$ indexes the network weights, and $m_j$ are their corresponding magnitudes such that $\mathbf{w}_{i,j} = m_j \mathbf{z}_{i,j}$.  
Using this representation, the integral in Eq.~\eqref{eq:prob4} can be approximated via Monte Carlo integration as
\begin{equation}\label{eq:prob3b}
    p(\alpha^{*}|\mathcal{G}^{*})\approx \frac{1}{N_d}\sum_{i=1}^{N_d} p(y^{*}|x^{*};\mathbf{w}_i),
\end{equation}
where $N_d$ denotes the number of neural networks sampled via dropout to form an ensemble.  
Practically, this is achieved by keeping dropout layers active with a given probability $p_{\mathrm{drop}}$ at inference time, enabling a Monte Carlo dropout procedure.  
This approach produces $N_d$ stochastic forward passes on the same input, allowing the estimation of statistical moments such as the mean and variance of predictions, and providing valuable insights into the reliability of the model’s outputs.

Specifically, the GNN $f(\cdot,\mathbf{w})$ has an architecture composed of a sequence of spectral graph convolutional layers (see \cite{kipf2017} for a detailed discussion) for updating the hidden states (features) $\mathbf{X}_l$ in layer $l$. This process is determined by the following propagation rule:
\begin{align}
    \mathbf{X}_{l+1} = \sigma\left( \mathbf{\bar{D}}^{-\frac{1}{2}}\mathbf{\bar{A}}\mathbf{\bar{D}}^{-\frac{1}{2}}\mathbf{X}_{l}\mathbf{W}_{l}\right)
\end{align}
where $\sigma$ is an activation function (i.e., hyperbolic tangent); $\mathbf{\bar{A}} = \mathbf{A} + \mathbf{I}$ denotes the adjacency matrix of the undirected graph $\mathcal{G}$; $\mathbf{\bar{D}}$ is the node degree matrix; and $\mathbf{W}_l$ is the trainable matrix of weights subjected to dropout operations for performing uncertainty quantification. 

\begin{figure}[!ht]
    \centering
    \includegraphics[width=\linewidth]{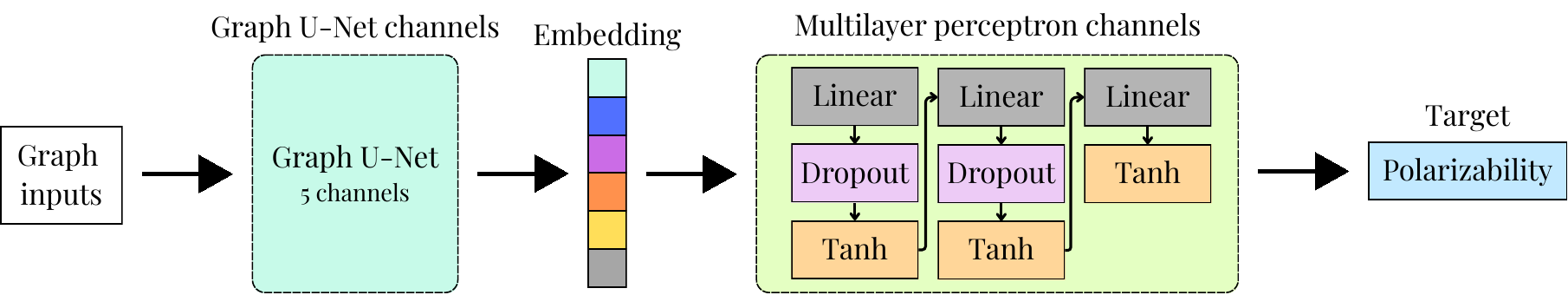}
    \caption{Graph U-Net model architecture.}
    \label{fig:graph_unet}
\end{figure}

The reference GNN predictive model implemented in \texttt{PolyGraphPy} consists of a graph-based U-Net architecture (see \cite{gao2019,ronneberger2015}), referred to as GraphUNet. This architecture enables effective graph embeddings that are subsequently processed by a sequence of MLP layers to predict the target property, such as the static polarizability $\alpha$. The model architecture, illustrated in Fig.~\ref{fig:graph_unet}, comprises GCN layers interleaved with top-$k$ pooling operations to downsample the graph while preserving the most informative nodes. The network is configured with five layers, a pooling ratio of 0.5, and summation-based skip connections to integrate multi-scale representations.

In the upsampling phase, the GraphUNet reconstructs the graph by restoring pooled nodes to their original positions using the indices recorded during pooling, effectively reversing the downsampling operation to recover spatial information. This reconstruction, combined with skip connections, enables the model to capture both local and global structural features that are critical for accurately predicting the properties of interest. Furthermore, a global mean pooling operation aggregates the node embeddings, which are then used as a polymer ``fingerprint'', which is used as input of an MLP network for property prediction.

\subsection{Property-guided molecular generation}
Generative models have emerged as powerful tools in computational chemistry for the \textit{de novo} design of molecules with targeted properties. Originally developed for natural language processing \cite{Vaswani2017, radford2019language}, generative pretrained transformers (GPT) have been adapted for molecular discovery by exploiting text-like representations of chemical structures. Trained on large molecular datasets, these models learn the syntactic and semantic rules underlying chemical representations, enabling them to generate novel and chemically valid structures \cite{Bilodeau2022, Frey2023}. However, some challenges persist, especially when related to the non-uniqueness of formats such as SMILES, which complicates the generation of consistently valid molecules.

As an alternative, genetic algorithms (GA), inspired by the concept of biological evolution, have long been used for molecular optimization \cite{Nigam2020, Jensen2019}, including polymers \cite{Kim2021ga}. GAs iteratively evolve a population of candidate molecules through selection, crossover, and mutation, guided by a fitness function that encodes the desired property (e.g., static polarizability). Although both GPT-based approaches and GAs have proven effective for generating molecules with tailored characteristics, they rely on fundamentally different search strategies: GPT models generate candidates probabilistically by sampling from a learned distribution, whereas GAs perform heuristic optimization driven by explicit evolutionary operations.

\begin{figure}[htbp]
    \centering
    \includegraphics[width=\linewidth]{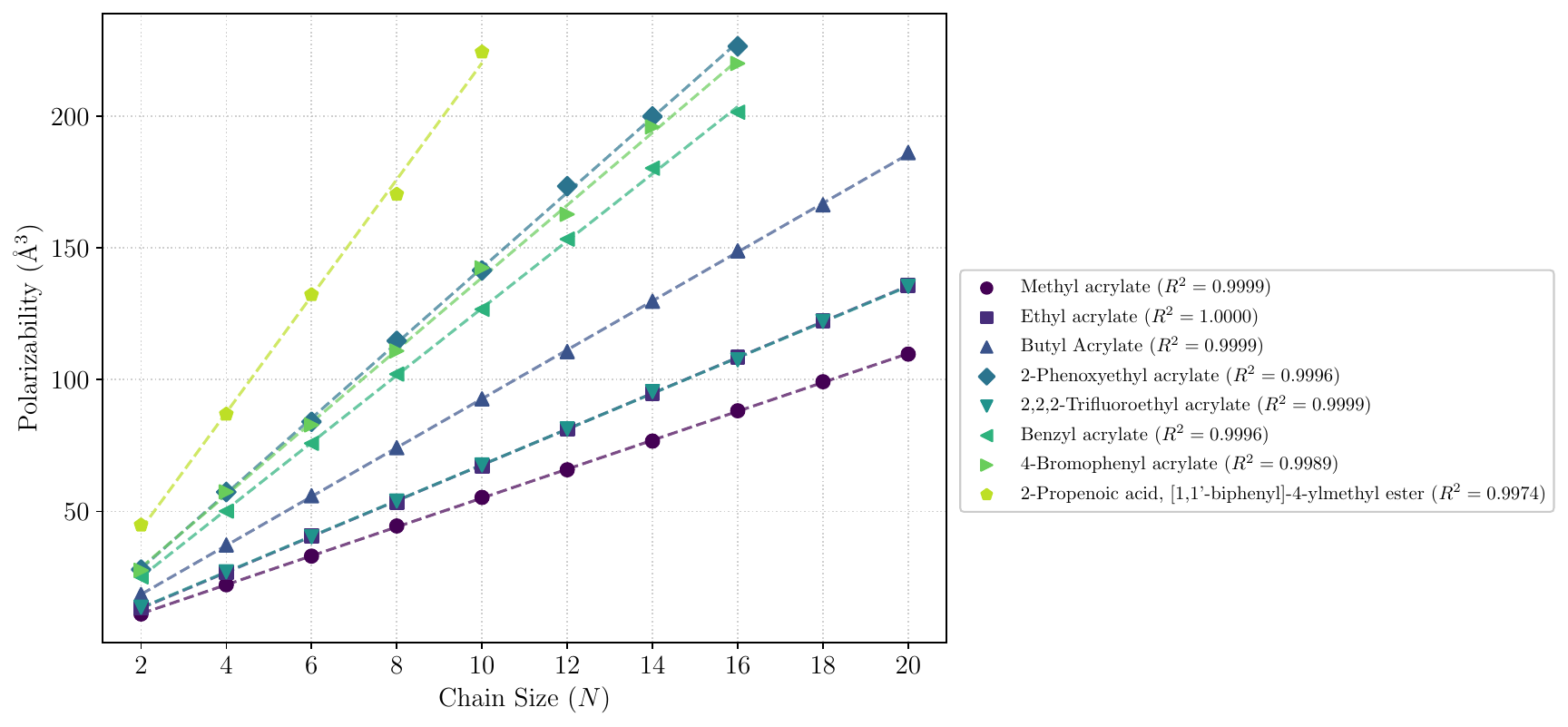}
    \caption{Scaling of static polarizability with respect to the degree of polymerization ($N$) for representative acrylate oligomers. All data points were computed using the automated DFTB+ quantum simulation pipeline integrated within \texttt{PolyGraphPy}. The strictly linear scaling behavior ($R^2 > 0.99$) quantitatively justifies the use of monomers as computational surrogates for homopolymer property optimization.}
    \label{fig:polarizability_vs_chain_size}
\end{figure}

In the current version of \texttt{PolyGraphPy}, we integrate these complementary generative paradigms to design monomers with improved static polarizability, using graph neural network (GNN) predictions to evaluate and filter generated structures. While the ultimate goal of the framework is polymer design (i.e., acrylates), monomers are employed herein as surrogates for acrylate homopolymers with targeted static polarizability. This is justified by the fact that a well defined functional relationship between the static polarizability and the chain size can be found. As illustrated in Fig.~\ref{fig:polarizability_vs_chain_size} for eight representative molecules selected at random and evaluated via the integrated DFTB+ pipeline, the relationship between the chain size ($N$) and the corresponding static polarizability is strictly linear across diverse acrylates ($R^2 > 0.99$), and this result is persistent even when other types of acrylates are considered. Consequently, optimizing the intrinsic polarizability of the repeating unit directly translates to an enhanced macroscopic response in the resulting homopolymer chain. Notably, this cannot be assumed for other uncorrelated properties, and a thorough investigation must be performed in such cases. Therefore, a potential generalization of this approach should take that into consideration.

The GPT-based workflow employs a transformer architecture trained on SELFIES-encoded monomers annotated with polarizability values, enabling sampling-based generation from the learned latent chemical space. In parallel, the GA-based workflow evolves a population of monomer candidates, with GNN-predicted polarizability serving as the fitness measure that directs the evolutionary search toward high-performing structures.

By combining generative modeling and predictive GNN frameworks, \texttt{PolyGraphPy} offers a unified platform for exploring and optimizing chemical spaces of polymers, thus facilitating systematic discovery of structures with enhanced target properties. Next, a detailed description of such models and their integration with \texttt{PolyGraphPy} is provided.

\subsubsection{Generative Pretrained Transformer}\label{sec:gpt}
The GPT-2 model has a decoder-only open source architecture and was released in 2019 \cite{radford2019language, HuggingFaceGPT2}. It serves as a foundation for the development of more recent large language models (LLMs) in the GPT family, including GPT-5. In this work, the small GPT-2 model is incorporated into \texttt{PolyGraphPy} as a generative engine trained on SELFIES representations of monomers. The GPT-2 model was selected primarily for its low VRAM footprint, which optimized resource utilization. In this context, the model contains 124 million parameters distributed across 12 transformer blocks, with a hidden size of 768, 12 attention heads, and a feed-forward network of width 3072. 

Within the \texttt{PolyGraphPy} framework, the LLM encodes molecular graphs into a latent space from which novel monomer structures can be sampled. The training procedure involves fine-tuning the pretrained GPT-2 model to prioritize monomers whose static polarizability values fall within a desired target range. These values are filtered using the predictive GNN models described earlier, retaining only monomers with the smallest discrepancy between the target polarizability and the GNN-predicted value. This fine-tuning step leverages GNN-predicted polarizability to guide the generative model toward chemically valid monomers aligned with the specified property.

The GPT model adopted here processes input sequences, such as SELFIES strings paired with polarizability values formatted as ``polarizability: X selfies: Y'' to generate molecular structures, where X denotes the desired polarizability target, and Y is the corresponding SELFIES pattern. Its core component is the multi-head self-attention mechanism, which computes attention scores for a token sequence $x = [x_1, x_2, \dots, x_n] $ as follows:
\begin{equation}\label{eq:att}
    \text{Attention}(\mathbf{Q}, \mathbf{K}, \mathbf{V}) = \text{softmax}\left( \frac{\mathbf{QK}^T}{\sqrt{d_k}} + \mathbf{M} \right)\mathbf{V},
\end{equation}
where $\mathbf{Q}=\mathbf{XW}_Q$, $\mathbf{K}=\mathbf{XW}_K$, and $ \mathbf{V}=\mathbf{XW}_V$ are the query, key, and value matrices, respectively, which are derived from the input embeddings $\mathbf{X}$ via learned weight matrices $\mathbf{W}_Q, \mathbf{W}_K, \mathbf{W}_V$. In addition, $d_k$ is the dimension of the key vectors and $\textbf{M}$ is a mask. The multi-head attention mechanism concatenates multiple attention outputs, allowing the model to capture diverse relationships within the sequence:
\begin{equation}
    \text{MultiHead}(\mathbf{X}) = \text{concat}(\text{head}_1, \dots, \text{head}_h)\mathbf{W}_O,
\end{equation}
where $\text{head}_i = \text{Attention}(\mathbf{XW}_{Q,i}, \mathbf{XW}_{K,i}, \mathbf{XW}_{V,i})$, $h$ is the number of heads, and $\mathbf{W}_O$ is a projection matrix. The transformer stacks multiple layers of multi-head attention, feed-forward networks, and layer normalization to produce a transition probability density function (PDF) over the next token in the sequence:

\begin{equation}\label{eq:gpt_pdf}
    \mathbf{P}(x_{t+1} | x_{1:t}) = \text{softmax}(\textbf{W}_{\text{out}}\textbf{h}_t),
\end{equation}
where $ h_t $ is the hidden state at position $ t $, and $ W_{\text{out}} $ maps it to the vocabulary size. During training, the model minimizes the cross-entropy loss over the SELFIES dataset:

\begin{equation}
    \mathcal{L} = -\sum_{t=1}^n \log \mathbf{P}(x_t | x_{1:t-1}, \theta),
\end{equation}
with $\theta$ representing the model parameters.

\begin{figure}[!ht]
    \centering
    \includegraphics[width=\linewidth]{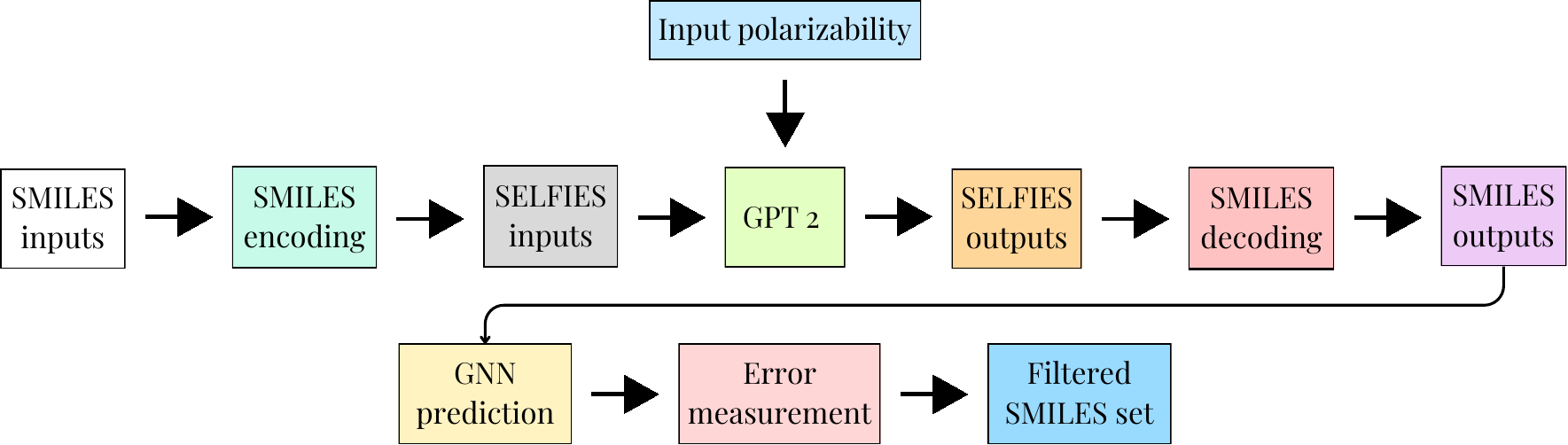}
    \caption{GPT model pretraining, training, and generation process.}
    \label{fig:gpt_process}
\end{figure}

Figure \ref{fig:gpt_process} illustrates the preprocessing, training, and generation pipeline for the GPT-based model. The process begins with SMILES strings from the molecular dataset, which are encoded into SELFIES representations using the \texttt{run} method from \texttt{GenerativePreprocess} class, as shown in the following code snippet: 
\begin{python}
prep = GenerativePreprocess(input_csv, generative_data_path)
prep.run()
\end{python}

The SELFIES strings, paired with normalized polarizability values, are used to fine-tune a pretrained GPT-2 model via the \texttt{GenerativeTrainer} class and its \texttt{run} method, as shown below: 
\begin{python}
trainer = GenerativeTrainer(generative_data_path, model_path, batch_size, learning_rate, epochs)
trainer.run()
\end{python}
During training, the model learns to associate SELFIES strings with target polarizability values, optimizing its parameters using the AdamW \cite{Loshchilov2019} optimizer and a mean squared error loss function. 

For property-guided generation, the \texttt{MoleculeGenerator} class takes the target polarizability values as input prompts, guiding the GPT-2 model to produce SELFIES strings, which are decoded into SMILES strings. This class is instantiated as follows:
\begin{python}
generator = MoleculeGenerator(model_path, output_path, monomers_number_per_target, threshold)
generator.run(targets)
\end{python}
The generated SMILES are verified for chemical validity using RDKit and selected based on structural properties, such as the presence of an acrylate substructure and the absence of multiple neutral fragments. The validated SMILES are converted into graph representations and passed through a pretrained GNN model to predict their static polarizability. Finally, a predetermined relative error threshold is applied to retain only monomers with a given error tolerance between the target and the predicted static polarizability $P_{static}$ and $\hat{P}_{static}$, respectively. Then, the accepted SMILES are stored as generated molecules. 

\subsubsection{Genetic algorithm}\label{sec:ga_methodology}
An alternative generative approach contained in the \texttt{PolyGraphPy} toolbox consists of the use of GA for generating monomers with targeted static polarizability. The GA iteratively optimizes a population of molecular structures through appropriate selection and crossover, guided by a fitness function that minimizes the deviation between predicted $\hat{P}_{static}$ and target static polarizability $P_{static}$. This approach is well-suited for exploring the chemical space of monomers, generating structures that align with the desired target value. 

\begin{figure}[!ht]
    \centering
    \includegraphics[width=\linewidth]{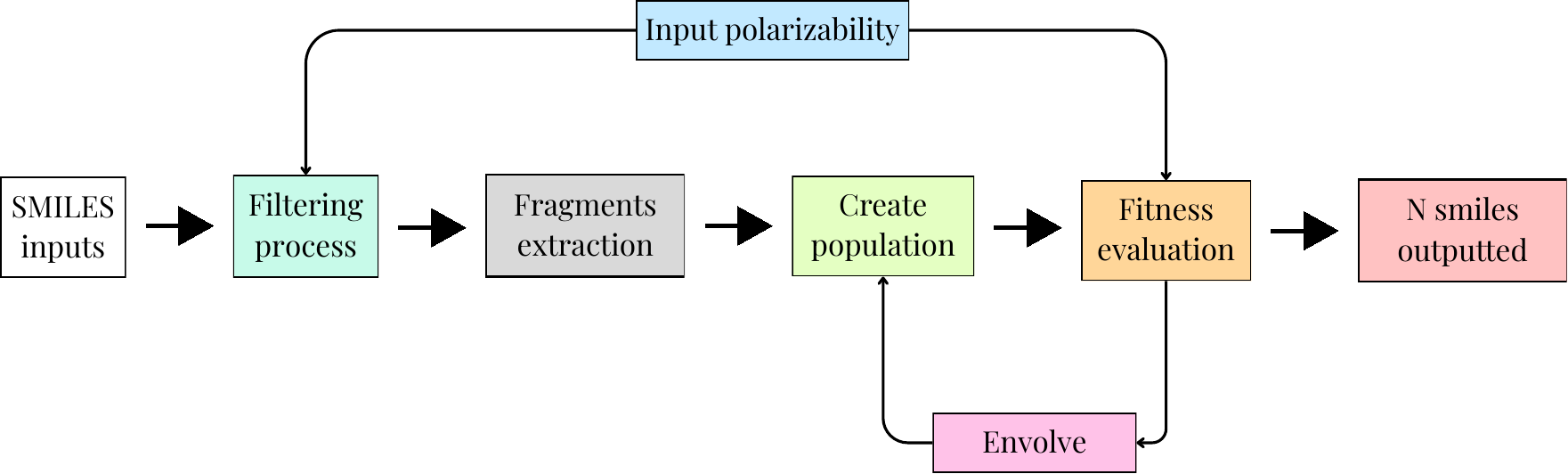}
    \caption{GA-based generative algorithm pipeline implemented in \texttt{PolyGraphPy}.t}
    \label{fig:ga_pipeline}
\end{figure}

The GA pipeline, which is depicted in Fig. \ref{fig:ga_pipeline}, begins by invoking the \texttt{GaModelLoader} class and its method \texttt{get\_components}. This is a loader tool that facilitates the preparation of all necessary components for the genetic algorithm, such as atoms and bond features, and loading the pretrained GNN model. This task is done in \texttt{PolyGraphPy} by using the following command:
\begin{python}
loader = GaModelLoader(input_csv, gnn_output_path, train_input_data_path, polymer_type, prediction_target)
model, preprocess, atom_encoder, bond_encoder = loader.get_components()
\end{python}
Next, the preprocessing is initialized by invoking the constructor of \texttt{FragmentGA} class, as shown below, to enable the appropriate selection of monomers with polarizability values within 90–110\% of the target and limit the number of atoms to at most 40 atoms. Moreover, molecular fragments are extracted using the BRICS algorithm and ensuring the presence of an acrylate backbone (\texttt{C=C-C(=O)O-[*]}).

\begin{python}
ga = FragmentGA(input_csv, model, preprocess, atom_encoder, bond_encoder, population_size, prediction_target, t)
fitness_scores = ga.run_parallel(generations, t)
\end{python}

The evolution process in the GA is executed with the function \texttt{run\_parallel} in class \texttt{FragmentGA}, and it consists of the following steps:
\begin{enumerate}
    \item Initialization: A population of $N = 100$ molecules is generated by the \texttt{build\_molecule} function as a default quantity, which randomly combines BRICS fragments to form chemically valid structures containing an acrylate backbone.
    \item Fitness evaluation: Each molecule is converted into a PyTorch Geometric \texttt{Data} object via the \texttt{\_mol\_to\_data} method. The pretrained GNN model predicts the static polarizability, and the fitness score is calculated as: $\text{fitness} = -|\hat{P}_{\text{static}} - P_{\text{static}}|$. Invalid molecules are assigned a fitness of $-1.0$.
    \item Selection: The population is sorted by fitness, and the top 50\% ($N/2 = 50$) of molecules are selected as parents for the next generation.
    \item Crossover: Fragments from the top-performing molecules are extracted using BRICS decomposition and recombined to form a new population of $N$ molecules, ensuring chemical validity and the presence of the acrylate backbone. Furthermore, 250 fragments are kept in the pool of fragments to balance computational overhead and maintain diversity.
    \item Iteration: The process iterates over 50 generations as a default value, and it consists of refining the population through repeated fitness evaluation, selection, and crossover.
\end{enumerate}

After this process, the best molecules (higher fitness) are accepted, and their SMILES are stored accordingly. 

\section{Results and Discussion}

\subsection{Input data: Acrylates}
\begin{figure}[!ht]
    \centering
    \includegraphics[width=\linewidth]{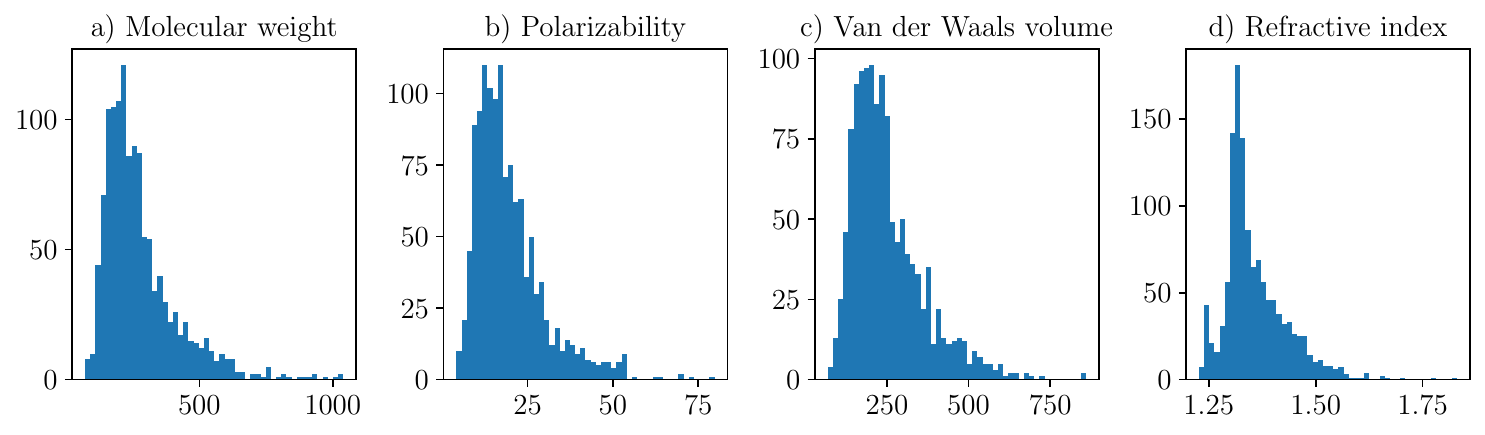}
    \caption{Histograms of numeric molecular descriptors found in the first dataset: (a) molecular weight computed using RDKit, (b) polarizability determined through the proposed atomistic simulation module, (c) Van der Waals volume, and (d) refractive index, with the latter two computed in the post-processing phase.}
    \label{fig:histogram_descriptors_original_data}
\end{figure}

Given the limited availability of publicly accessible datasets for monomers and polymers with computed polarizability values, two datasets were generated via atomistic simulations using DFTB+. The first dataset comprises 1,264 acrylate monomers with corresponding SMILES extracted from PubChem \cite{Kim2021, pubchem_cite}, which includes compounds such as methyl acrylate, butyl acrylate, ethyl acrylate, and acrylic acid, each characterized by the \texttt{C=C-C(=O)O} substructure critical for polymer applications \cite{Young1999}. Fig. \ref{fig:histogram_descriptors_original_data} shows the some molecular descriptors showin the diversity of this dataset. 

\begin{figure}[!ht]
    \centering
    \includegraphics[width=\linewidth]{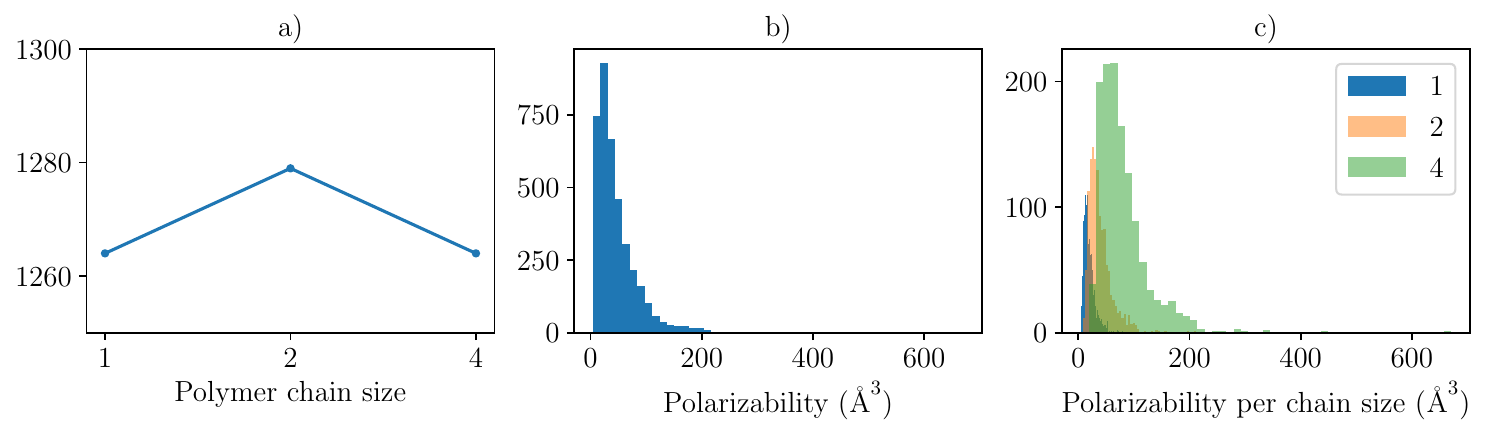}
    \caption{Plot showing the distribution of polymers per chain size (a), alongside histograms of static polarizability for the entire dataset (b) and stratified by chain size (c).}
    \label{fig:histogram_polarizability_homopolymer}
\end{figure}

The monomers from the first dataset were used to construct a second dataset of monomers and homopolymers, where 3,808 DFTB atomistic simulations were performed to compute the static polarizability $\alpha$. For this dataset, the chain size was varied to include molecules with 1 monomer, 2 monomers, and 4 monomers. The simulated polarizability and chain size distributions are plotted in Fig. \ref{fig:histogram_polarizability_homopolymer}, and it reveals a clear trend in Fig. \ref{fig:histogram_polarizability_homopolymer}c between the chain size and $\alpha$. This can be attributed to the approximately linear relationship between the number of atoms in a molecule and its polarizability \cite{Young1999}.

Additionally, a third dataset focusing on copolymers was also constructed with the monomers from the first dataset. Initially, to reduce computational overhead and limit the chemical space, only monomers with 25 or fewer atoms were retained. Furthermore, molecules containing metallic elements were excluded to ensure compatibility with the DFTB+ parameters. The resulting data were then partitioned into 16 clusters using the Recursive Coordinate Bisection (RCB) method, based on a 3D feature space comprising molecular weight, complexity, and polar surface area, all computed using RDKit. To ensure structural diversity, 9 monomers were randomly sampled from each of the 16 clusters, resulting in a diverse pool of $n = 144$ representative monomers. Every monomer in this pool was then paired with every other monomer, yielding $\frac{n(n-1)}{2} = 10,296$ unique copolymer combinations. These copolymer pairs were subsequently simulated using DFTB+ to compute their static polarizability, resulting in 8,627 successfully completed atomistic simulations with \texttt{chain\_size} set to 1 (representing a single A-B copolymer unit). 

\begin{figure}[!ht]
    \centering
    \includegraphics[width=0.55\linewidth]{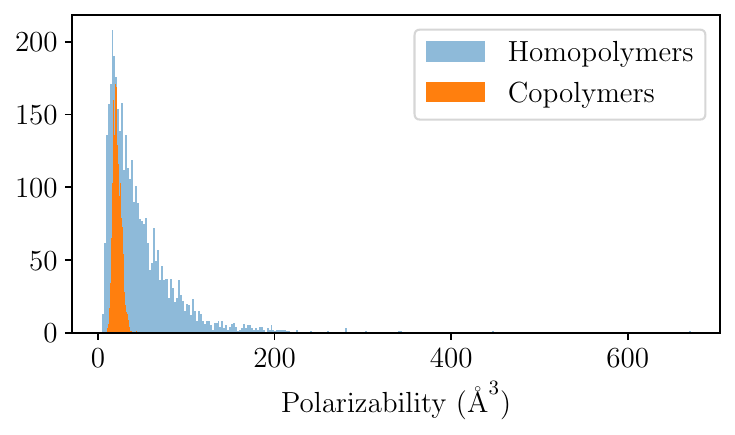}
    \caption{Distributions of static polarizability for the copolymer and homopolymer/monomer datasets.}
    \label{fig:copoly_polarizability}
\end{figure}

The polarizability distribution, illustrated in Figure \ref{fig:copoly_polarizability}, shows a high concentration of the polarizability values for copolymers in the interval $[9.65, 42.91]$, contrasting with the broader distribution observed in the homopolymer dataset. This concentration is primarily due to the exclusion of monomers with more than 25 atoms.

\subsection{Polarizability prediction}
Following the construction of the datasets, two predictive models were trained using the Graph U-Net architecture (illustrated in Figure \ref{fig:graph_unet}) to predict static polarizability. One model was trained specifically on the monomer/homopolymer dataset, while the second was trained on the copolymer dataset. The monomer/homopolymer dataset, comprising 3,808 graphs, was partitioned into 3,427 molecular graphs for training and 381 for validation (a 90/10 split). Similarly, the copolymer dataset, totaling 8,627 graphs, was divided into 7,764 graphs for training and 863 for validation.

Both models demonstrated strong predictive capabilities and good generalization on their respective validation sets. The GNN model trained on the monomer/homopolymer data achieved $\text{MAPE} = 11.83\%$, $R^2 = 0.9739$, and $\text{MSE} = 0.0015$. The copolymer GNN model exhibited even lower error margins, achieving $\text{MAPE} = 5.19\%$, $R^2 = 0.9745$, and $\text{MSE} = 0.00093$.

\begin{figure}[!ht]
    \centering
    \includegraphics[width=\linewidth]{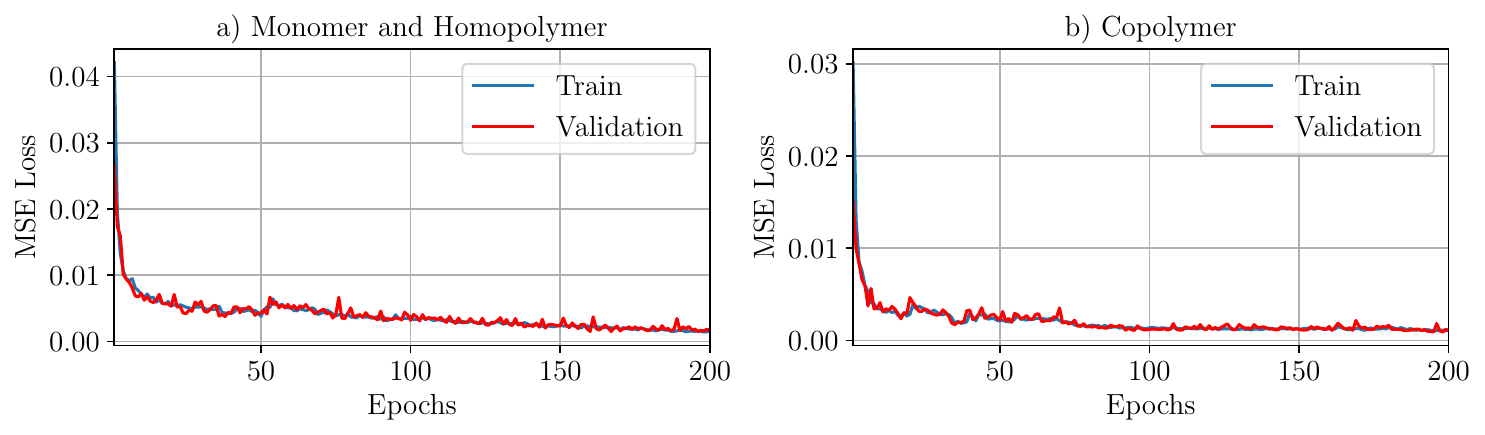}
    \caption{Training and validation losses for (a) the monomer and homopolymer dataset, and (b) the copolymer dataset.}
    \label{fig:training_val_losses}
\end{figure}

\begin{figure}[!ht]
    \centering
    \includegraphics[width=\linewidth]{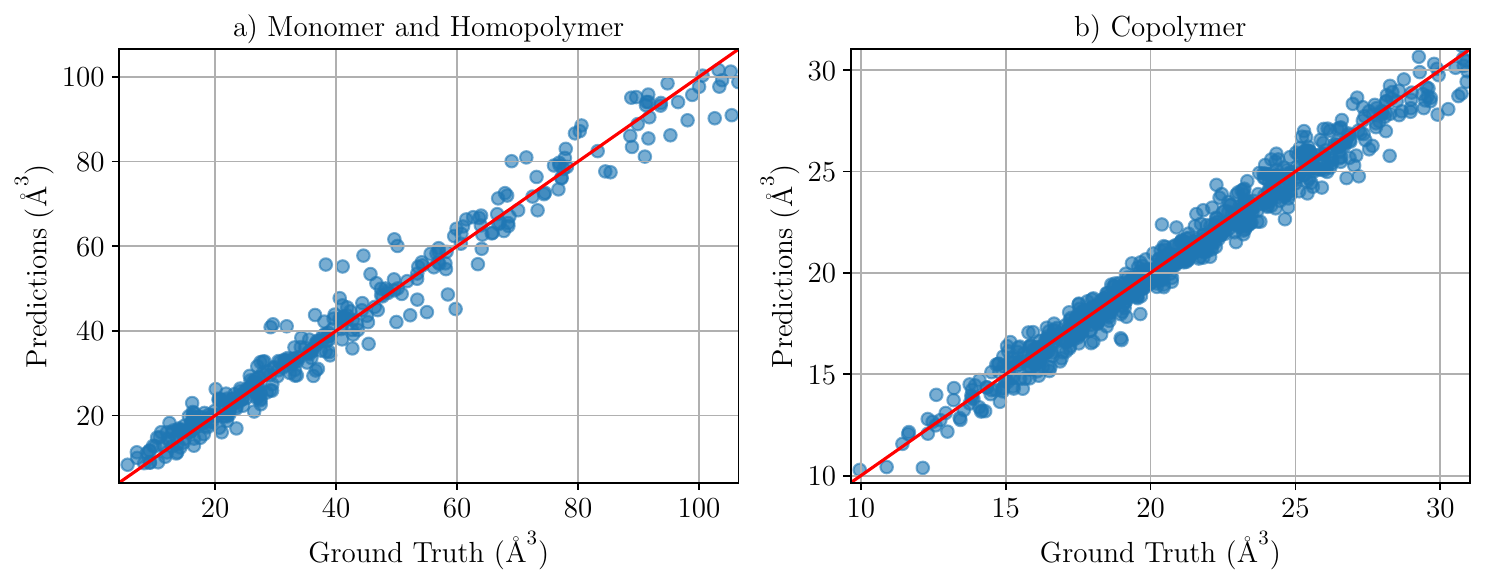}
    \caption{Predicted versus ground truth polarizability for the validation sets of (a) the monomer and homopolymer dataset, and (b) the copolymer dataset.}
    \label{fig:t_vs_p}
\end{figure}

Figure \ref{fig:training_val_losses} shows the training and validation losses for both GNN models. Additionally, Figure \ref{fig:t_vs_p} illustrates the predicted versus ground truth polarizability plots for the validation sets, further reinforcing the robustness of both models. 

Furthermore, both models employ dropout regularization to mitigate overfitting. in this context, uncertainty quantification was conducted to assess prediction variability, thereby enhancing the overall reliability of the calculated static polarizabilities. This is achieved by interpreting standard dropout as approximate variational inference in a deep Gaussian process, which enables the computation of a predictive mean and variance through multiple stochastic forward passes during inference \cite{Gal2016, Gal2016appendix}. 

\begin{figure}[!t]
    \centering
    \includegraphics[width=\linewidth]{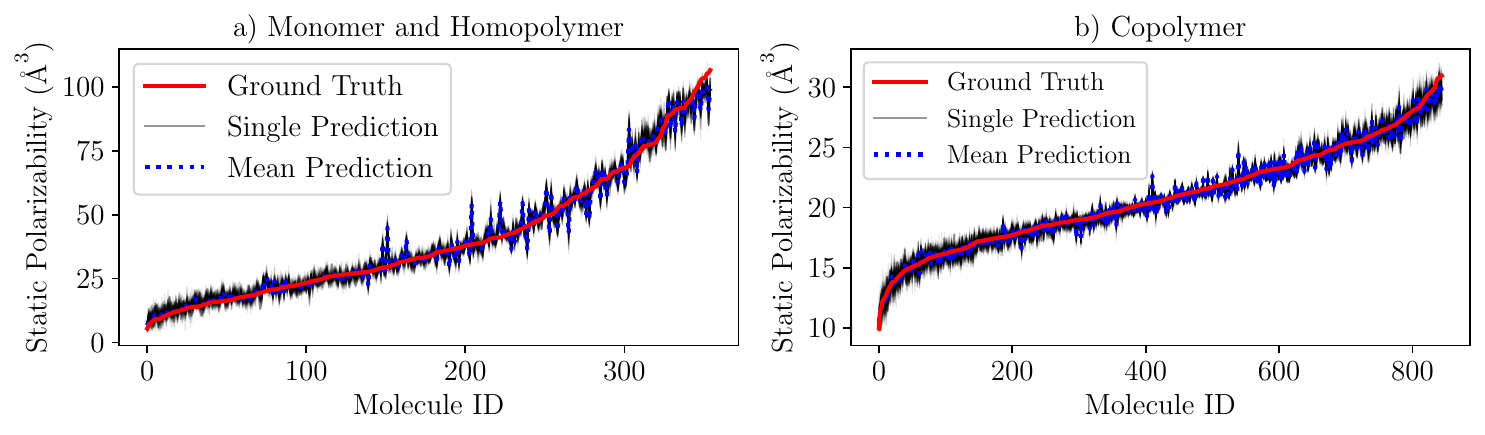}
    \caption{Comparison of ground truth and predicted static polarizability from 100 Monte Carlo runs with dropout enabled, showing controlled variance and close alignment of mean predictions for (a) the monomer/homopolymer dataset and (b) the copolymer validation dataset.}
    \label{fig:monte_carlo_y_pred}
\end{figure}

The comparison between ground truth polarizability and predictions from 100 Monte Carlo (MC) runs with dropout enabled is shown in Figure \ref{fig:monte_carlo_y_pred}. The GNN models produce varying predictions across runs, yet the variance remains well-controlled, indicating robust model performance. Moreover, the mean predictions closely align with the ground truth values for both the monomer/homopolymer and copolymer validation datasets, reinforcing the reliability of the proposed computational framework.

\begin{figure}[!ht]
    \centering
    \includegraphics[width=\linewidth]{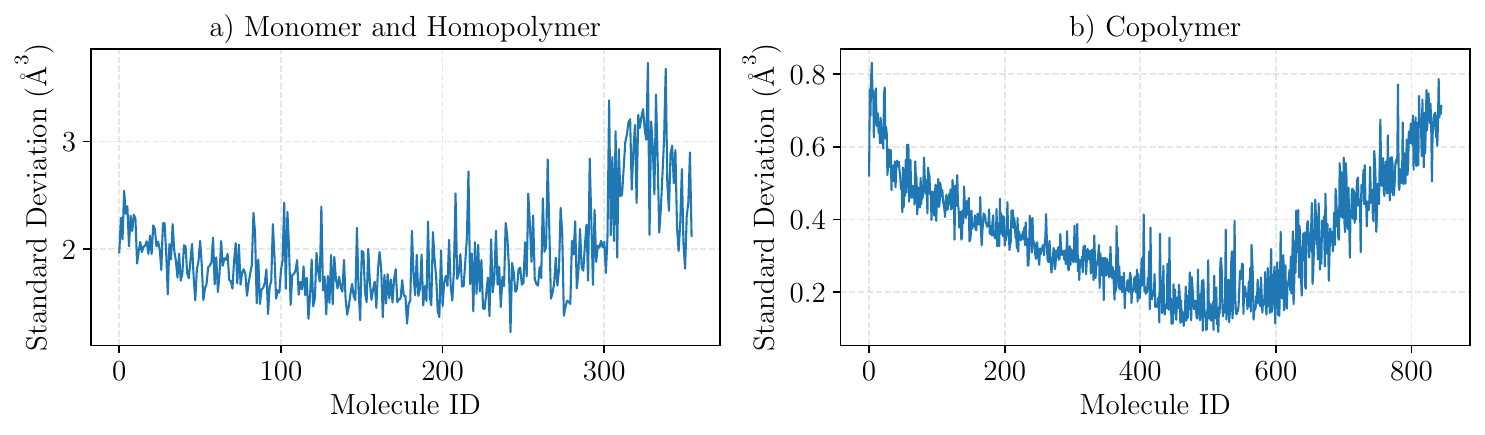}
    \caption{Scaled polarizability standard deviation (STD) per molecule from 100 Monte Carlo runs with dropout enabled during validation, showing the STD for (a) the monomer/homopolymer dataset and (b) the copolymer dataset.}
    \label{fig:std}
\end{figure}

Additionally, Figure \ref{fig:std} shows the polarizability standard deviation (STD) per molecule, derived from 100 MC runs with dropout activated during the validation phase. The STD remains below 0.9 for copolymers and below 3.8 for monomers/homopolymers. This discrepancy in uncertainty can be directly attributed to the amount of data used to train each model: the substantially larger copolymer dataset provides more training evidence, which correspondingly reduces the model's epistemic uncertainty compared to the smaller monomer/homopolymer dataset.

\begin{figure}[!t]
    \centering
    \includegraphics[width=\linewidth]{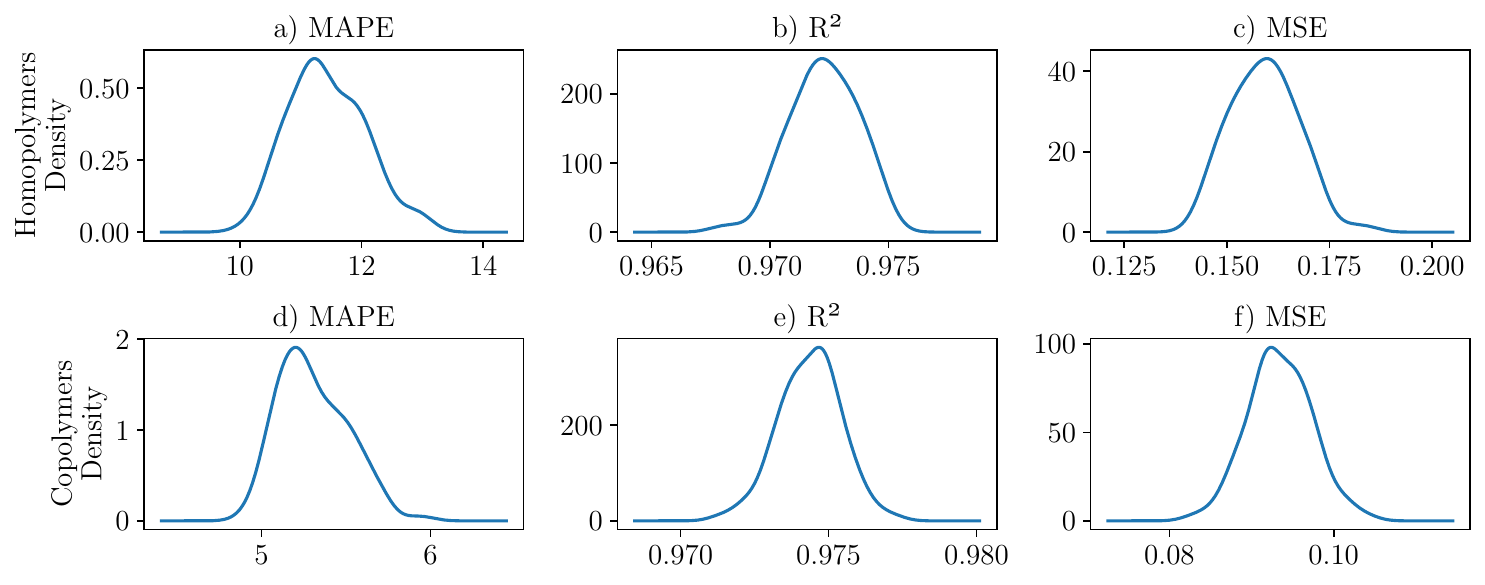}
    \caption{Estimated PDFs of MAPE, R$^2$, and MSE for 100 validation runs, showing concentrated means and near-Gaussian behavior for monomers/homopolymers (a–c) and copolymers (d–f).}
    \label{fig:pdfs_errors}
\end{figure}

Figure \ref{fig:pdfs_errors} presents the PDFs of the MAPE, R$^2$, and MSE metrics computed for the validation datasets over 100 runs, for monomers/homopolymers (panels a–c) and copolymers (panels d–f). The PDFs in panels a–c exhibit a high concentration around their respective means with short tails, closely approximating Gaussian distributions. A similar behavior is observed for the PDFs shown in panels d–f.

\subsection{Polarizability-guided high-throughput generation}
Following the training and validation of the GNN models, we proceeded to evaluate the two generative models implemented in \texttt{PolyGraphPy}: the GA-based and the GPT-based models. To assess the robustness of the GA-based generative model, the following strategy was adopted:

\begin{enumerate}
    \item A set of 100 target polarizability values was generated using \texttt{np.linspace(P01, P75, 100)}, where 
    \begin{itemize}
    \item  \texttt{P01 = df[`target\_scaled'].quantile(0.01)}
    \item \texttt{P75 = df[`target\_scaled'].quantile(0.75)}
    \end{itemize}
    This interval was selected to optimize computational efficiency while covering a considerable range of scaled polarizability values, as initial tests revealed that the BRICS algorithm exhibits reduced performance for molecules with 40 or more atoms.
    \item For each target value $i$, the GA-based model was executed following the methodology detailed in the previous sections.
    \item For each target $i$, 100 candidate monomers were generated based on the defined population size, yielding a total of 10,000 raw candidate monomers.
\end{enumerate}

\begin{figure}[!ht]
    \centering
    \includegraphics[width=\linewidth]{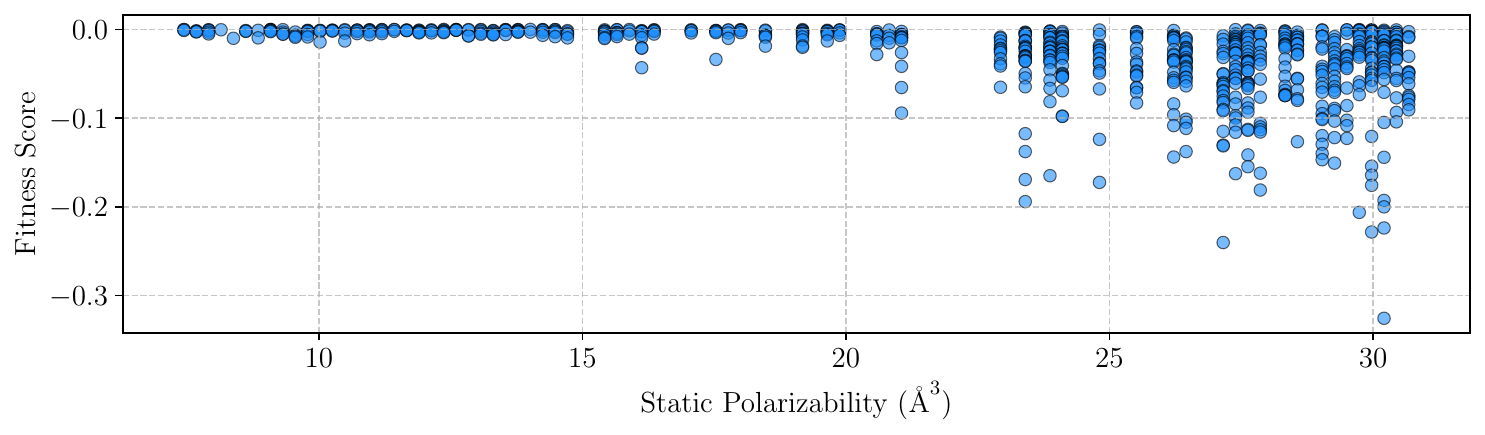}
    \caption{Scatter plot showing the fitness score as a function of the static polarizability for the generated monomers.}
    \label{fig:waterfall}
\end{figure}

Following the initial generation phase, we evaluated the performance and yield of the GA-based model. Out of the 10,000 candidates, 730 valid and unique SMILES strings were successfully generated. Notably, 89.99\% of these were entirely absent from the original acrylate monomer dataset, highlighting the strong exploratory and generative capacity of the GA model. 

\begin{figure}[!ht]
    \centering
    \includegraphics[width=\linewidth]{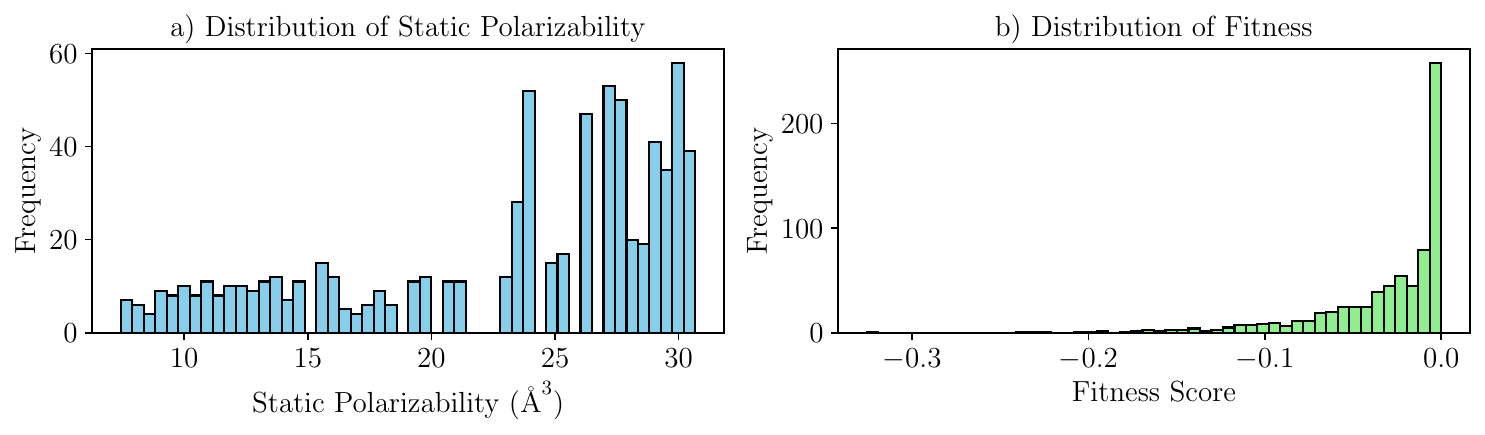}
    \caption{Distributions of (a) static polarizability and (b) fitness scores for the valid generated monomers.}
    \label{fig:ga_robustness}
\end{figure}

Figure~\ref{fig:waterfall} illustrates the distribution of the corresponding fitness scores as a function of static polarizability, indicating that the model successfully generates acrylate monomers with very low error for polarizability values $\leq 20$ \AA{}$^3$. Additionally, Figure \ref{fig:ga_robustness}a summarizes the frequency of generated monomers by static polarizability, showing that the genetic algorithm approach produces a higher volume of valid and unique acrylates at $\geq 22$ \AA{}$^3$. Finally, Figure \ref{fig:ga_robustness}b displays the distribution of the fitness scores, revealing a sharp concentration around $\approx 0$. This demonstrates the high reliability and precision of the GA algorithm when operating in tandem with the predictive GNN model.

\begin{figure}[!h]
    \centering
    \includegraphics[width=0.5\linewidth]{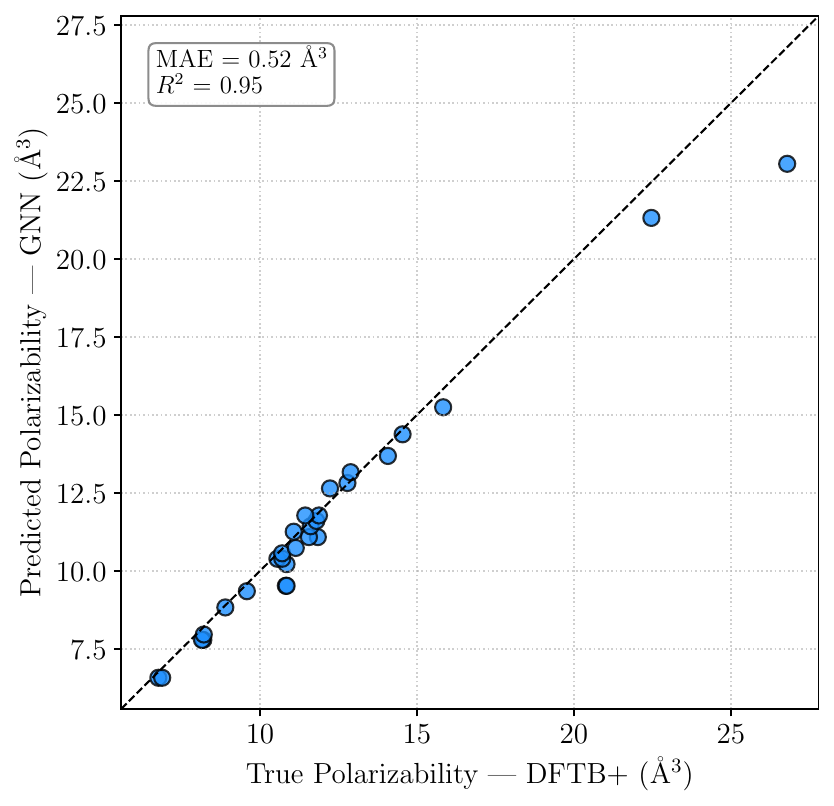}
    \caption{Parity plot comparing the GNN-predicted static polarizability (surrogate fitness) against the ground-truth DFTB+ calculated values for the top 30 novel monomers generated by the GA. The strong correlation ($R^2 = 0.95$, $\text{MAE} = 0.52$ \AA{}$^3$) independently validates the reliability of the generative pipeline.}
    \label{fig:dftb_vs_gnn_parity}
\end{figure}

To further validate the generative framework while ensuring that the GNN surrogate accurately captures the properties of novel molecular architectures, the top 30 newly generated monomers were selected and their SMILES codes were subjected to independent DFTB+ atomistic simulations. The static polarizabilities calculated by the DFTB+ pipeline were then compared directly against the original predictions yielded by the GNN surrogate. As illustrated in Fig.~\ref{fig:dftb_vs_gnn_parity}, the GNN demonstrated exceptional transferability to these newly generated structures, achieving a Mean Absolute Error (MAE) of 0.52 \AA{}$^3$ and a coefficient of determination ($R^2$) of 0.95. This independent physical verification confirms that the GA reliably discovers novel chemistries with true, physically verifiable target properties, effectively ruling out the generation of adversarial artifacts that merely exploit the predictive model, while maintaining the necessary chemistry constraints.

To assess the robustness and reliability of the GPT-based generative model, the following strategy was adopted:

\begin{enumerate}
    \item A set of 200 target polarizability values was created using \texttt{np.linspace(0, 1, 200)} to thoroughly explore the scaled polarizability range.
    \item For each target value $i$, the GPT-based model was executed following the methodology detailed in previous sections, generating five SELFIES strings per polarizability target, yielding a total of 1,000 raw generations.
    \item For each generation, RDKit was utilized to validate the SMILES strings decoded from the SELFIES, ensuring both chemical validity and the presence of the required \texttt{C=C-C(=O)O} substructure. This filtering process resulted in a final dataset comprising 126 valid monomers.
\end{enumerate}

The performance of the GPT-based model was evaluated in parallel with the GA-based model. Out of the 126 valid monomers, 99.2\% were unique (125 distinct SMILES strings), demonstrating high generational diversity. Moreover, 78.23\% of these valid, generated SMILES were entirely absent from the original acrylate monomer training dataset, highlighting the model’s strong capability for exploring novel chemical space. 

While a 12.6\% yield (126 valid structures from 1,000 raw strings) may initially appear limited, it is essential to contextualize this success rate within the highly constrained design space of targeted polymer building blocks. To establish a quantitative baseline, an experiment utilizing random token sampling from the SELFIES vocabulary to generate 1,000 strings of comparable lengths was performed. This random baseline yielded a 0\% success rate for producing valid molecules that satisfied the strict \texttt{C=C-C(=O)O} acrylate substructure constraint. 

Therefore, the GPT model provides a massive probabilistic enrichment over random chemical exploration. Furthermore, the high novelty rate (78.23\%) strongly suggests that the model is not overfitting or merely memorizing the training dataset, but is actively recombining learned latent representations into novel chemistries. An in-depth analysis of the 87.4\% rejected strings revealed two primary failure modes. First, because SELFIES mathematically guarantees generic chemical valency, the SELFIES-to-SMILES conversion itself is rarely a bottleneck for basic validity. Instead, the vast majority of failures occurred because the generated molecules lacked the mandatory acrylate backbone, thus failing the domain-specific RDKit filter. Second, sequence truncation occasionally occurred, where the autoregressive model reached maximum token limits before properly closing complex side-chain branches. Despite these failure modes, the practical utility of the GPT model remains remarkably high: because the automated RDKit filtering and sub-millisecond GNN proxy operate in seconds, the 87.4\% of invalid strings are discarded with negligible computational overhead, effectively functioning as an ultra-fast molecular generator.

\begin{figure}[!ht]
    \centering
    \includegraphics[width=\linewidth]{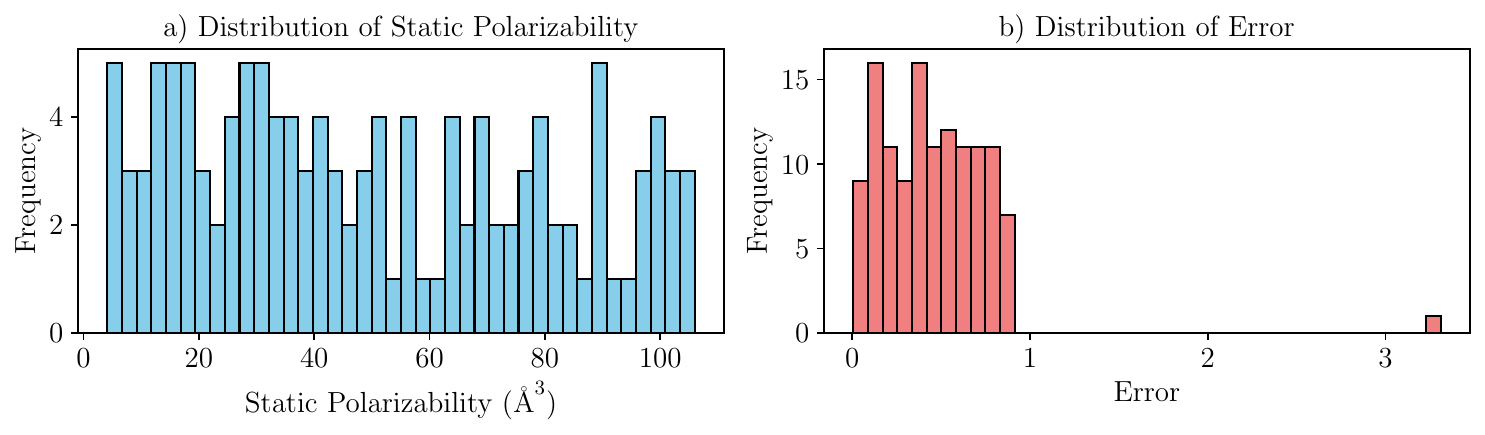}
    \caption{Distributions of (a) static polarizability and (b) relative error for the valid monomers generated by the GPT-based model.}
    \label{fig:results_gpt}
\end{figure}

In addition, Fig. \ref{fig:results_gpt} presents the results obtained with the GPT-based approach, where panel (a) illustrates the distribution of generated monomers by their static polarizability, while panel (b) displays the relative error distribution. Although the GPT-based approach is capable of generating valid monomers more uniformly and exploring a broader range of static polarizability values than the GA approach, its relative errors are noticeably larger in comparison with the GA-based approach.

As a highly complex probabilistic framework, the GPT-based model struggles to precisely match specific target polarizability values, an issue amplified by the relatively small size of the training dataset. Furthermore, the generative performance is intrinsically tied to the computational resources utilized in this study. The current GPT-2 model was trained and deployed on an NVIDIA RTX A2000 GPU, which imposes a strict hardware limitation of roughly 6 GB of Video RAM (VRAM). This memory constraint inherently restricts the maximum number of trainable parameters, the context window size, and the batch sizes during training. One can anticipate that transitioning to more powerful hardware with higher VRAM capacities would allow for the deployment of more advanced, larger-scale transformer architectures. Such an upgrade would naturally enhance the model's syntax resolution capabilities, yielding a higher percentage of valid structures and significantly improving its precision in targeting specific polarizability ranges. To address this limitation in future iterations, a potential strategy involves expanding the training dataset to include homopolymers and copolymers. By treating these larger structures as ``super monomers'' within the generative framework, the model could achieve better coverage and accuracy across higher polarizability ranges.

\subsection{Polymer space exploration}
\begin{figure}[!h]
    \centering
    \includegraphics[width=\linewidth]{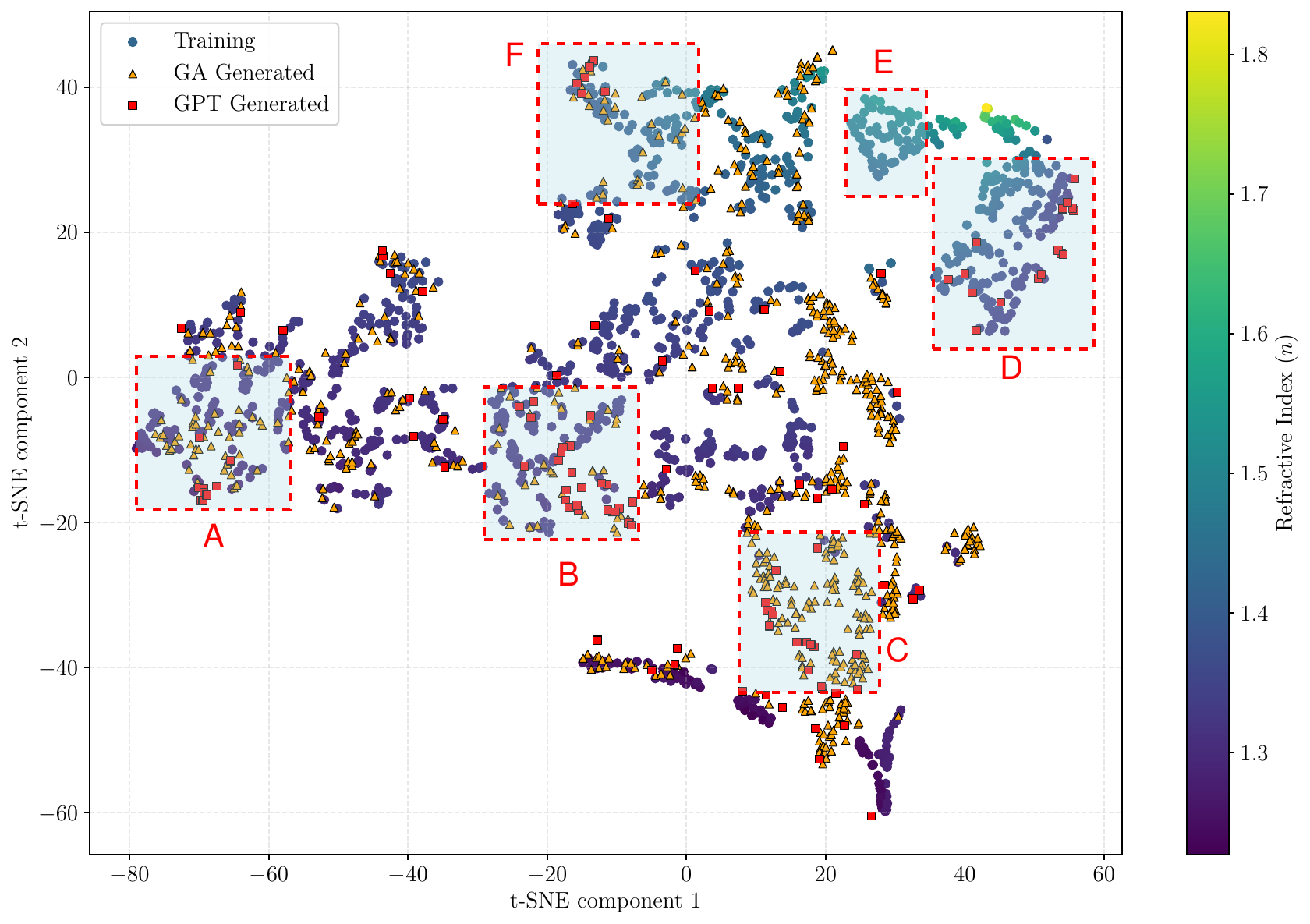}
    \caption{t-SNE visualization of the chemical space comparing the original acrylate dataset with the valid monomers generated by the GA-based and GPT-based models. Six areas are also selected to assess the adeherence of the generative models with the training dataset as well as their exploratory capabilities.}
    \label{fig:ga_gpt_map}
\end{figure}

Figure \ref{fig:ga_gpt_map} uses t-distributed Stochastic Neighbor Embedding (t-SNE \cite{Maaten2008}) to visualize the chemical space and identify where the generative approaches effectively explore new monomers. This 2D embedding was constructed using four key physical properties: (i) molecular weight (MW), (ii) static polarizability $\alpha$, (iii) van der Waals (vdW) volume, and (iv) refractive index $n$. Note that the relation between the refractive index and the static polarizability is determined by the Lorentz-Lorenz formula  \cite{lorenz1869experimentelle,lorentz1880beziehung,born1999principles}, which is given by
\begin{align}
    \frac{n^2-1}{n^2+2} = \frac{4\pi}{3}N\alpha
\end{align}
where $N$ is the number density of molecules (molecules per unit volume).

This projection enables effective visualization of the complex chemical space, facilitating the assessment of the generative methods' exploratory capabilities. Notably, both the GA- and GPT-based methods exhibit distinct exploratory behaviors while maintaining strong adherence to the underlying polymer space structure defined by the training data (see Fig.~\ref{fig:ga_gpt_map}). Overall, while the GA model explores the broader polymer space more effectively, the GPT model reaches specific regions not fully exploited by the GA.

Locally, however, each model's performance varies based on the structural features and physical properties of the generated molecules. To investigate this, six areas of interest are selected (see Fig.~\ref{fig:ga_gpt_map}) that correspond to distinct property clusters. Because the training dataset encompasses a wide range of chemical motifs---including fluorocarbons, complex sugars, and dense aromatics---the specific characteristics of the molecules within each selected region are detailed in Table~\ref{tab:training}.

\begin{table}[htbp]
    \centering
    \caption{Qualitative characteristics for the molecules in each highlighted area of Fig.~\ref{fig:ga_gpt_map}}
    \renewcommand{\arraystretch}{1.5}
    \begin{tabularx}{\textwidth}{@{} p{0.1\textwidth} p{0.3\textwidth} X @{}}
        \toprule
        \textbf{Area} & \textbf{Characteristics} & \textbf{Description} \\
        \midrule
        \textbf{A} & Low refractive index, and low MW and vdW volume  & Dominated by small, highly fluorinated side chains. They achieve low refractive indexes (e.g., around 1.2 - 1.3) through low molecular weights (e.g., around 80 - 160) and tightly packed structures. \\ 
        \textbf{B} & Moderate vdW volume, complex oxygenation & Low refractive indexes (e.g., around 1.2 - 1.3) (but overall a bit higher than in area A), characterized by complex, oxygen-rich side chains---polyols, multi-ester linkages, and short polyethylene glycol (PEG) components. \\
        \textbf{C} & Higher vdW volume, higher polarizability and refractive index & Achieves higher MW and vdW volume through highly branched, complex oxygen networks, including carbohydrate-like ring structures and extended ether linkages. \\
        \textbf{D} & Longer molecules with higher vdW volume, polarizability, and refractive index & Composed of, for example, of long-chain aliphatic systems, bulky spacers and protected macromonomers. \\
        \textbf{E} & Dense aromatics with higher refractive index ($>1.4$) & Highly conjugated systems featuring multiple fused aromatic rings, azo bonds ($-\mathrm{N}=\mathrm{N}-$), and sulfones. \\
        \textbf{F} & Small aromatics and cyclics & Moderate refractive index (around 1.37), simple substituted phenyl rings and small cycloaliphatics. \\
        \bottomrule
    \end{tabularx}\label{tab:training}
\end{table}

\begin{figure}[!h]
    \centering
    \includegraphics[width=\linewidth]{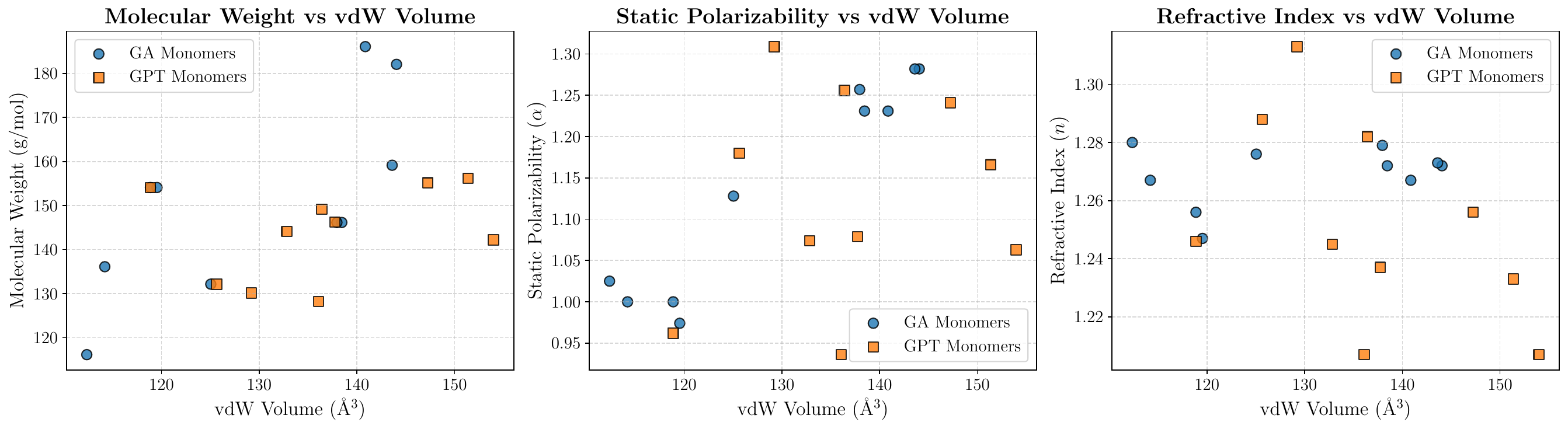}
    \caption{Comparison of vdW volume and (a) molecular weight, (b) static polarizability, and (c) refractive index for GA- and GPT-based models in Area A.}
    \label{fig:area_A}
\end{figure}

The performance of each generative model within each selected area is evaluated based on its ten best-generated monomers, where applicable (see the detailed data in the supplementary material). Additionally, plots correlating vdW volume against molecular weight, static polarizability, and refractive index are provided for each area. In \emph{Area A}, the GA model generates new molecules that demonstrate strong adherence to the fluorinated chain motifs found in the training data. This exploitation is achieved through appropriate permutations of the substituents in the molecules' R groups. In contrast, the GPT model explores branched aliphatics and simple heterocycles. As shown in Fig.~\ref{fig:area_A}, both models generate molecules with vdW volumes between $110~\AA^3$ and $160~\AA^3$, MW between $110~\mathrm{g/mol}$ and $190~\mathrm{g/mol}$, static polarizabilities between $0.9$ and $1.35$, and low refractive indices between $1.2$ and $1.32$. However, the GPT model spans a broader range of refractive indices compared to the GA model.

\begin{figure}[!h]
    \centering
    \includegraphics[width=\linewidth]{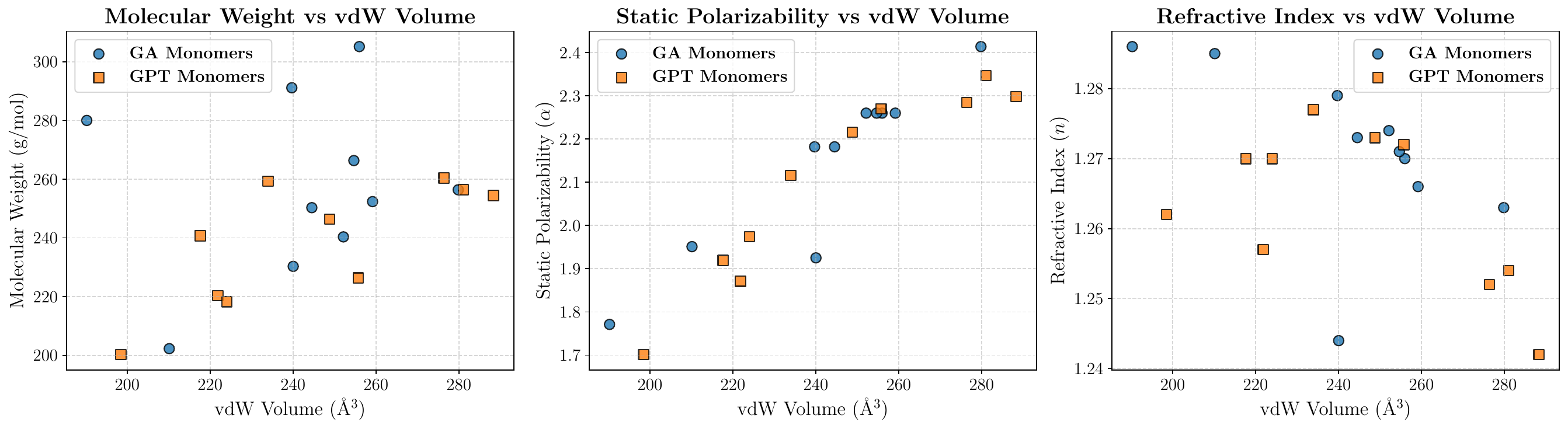}
    \caption{Comparison of vdW volume and (a) molecular weight, (b) static polarizability, and (c) refractive index for GA- and GPT-based models in Area B.}
    \label{fig:area_B}
\end{figure}

Among the top ten monomers in \emph{Area B}, the GA model deviates from the oxygen-heavy training data. It relies on bulky aliphatic bridged rings and halogens to achieve the target properties, whereas the GPT-based model shows a preference for linear geometries, leaning heavily on straight alkyl chains alongside simple aromatics. As shown in Fig. \ref{fig:area_B}, both models generate molecules with higher vdW volumes ($200$ to $300~\AA^3$), higher molecular weights (200 to 310 g/mol), and higher static polarizabilities (1.7 to 2.5). However, the generated refractive indices fall within a narrower range (1.24 to 1.30) compared to those obtained for the molecules in Area A.

\begin{figure}[!h]
    \centering
    \includegraphics[width=\linewidth]{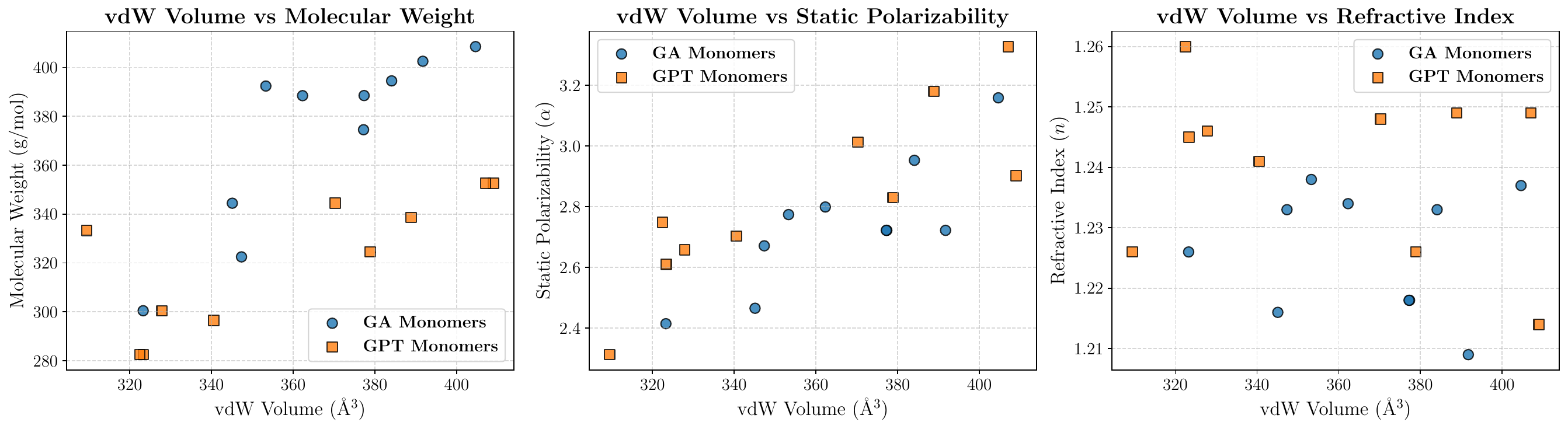}
    \caption{Comparison of vdW volume and (a) molecular weight, (b) static polarizability, and (c) refractive index for GA- and GPT-based models in Area C.}
    \label{fig:area_C}
\end{figure}

In \emph{Area C}, the training data achieves high mass and volume through highly branched, complex oxygen networks. While the GA model achieves the target properties by incorporating bulky polycyclic aliphatics, bridged rings, and aromatic ethers—resulting in high topological complexity, the GPT model focuses on generating extremely long linear or lightly branched hydrocarbon chains. As shown in Fig.~\ref{fig:area_C}, both models generate molecules with higher vdW volumes (between $300~\AA^3$ and $420~\AA^3$), higher MW (between $280~\mathrm{g/mol}$ and $420~\mathrm{g/mol}$), higher static polarizabilities ($2.2$ to $3.4$), and relatively lower refractive indices ($1.2$ to $1.26$). Notably, the amount of training data in this area is small, demonstrating that even under such conditions, the generative models are able to successfully fill gaps in the latent space.

\begin{figure}[!h]
    \centering
    \includegraphics[width=\linewidth]{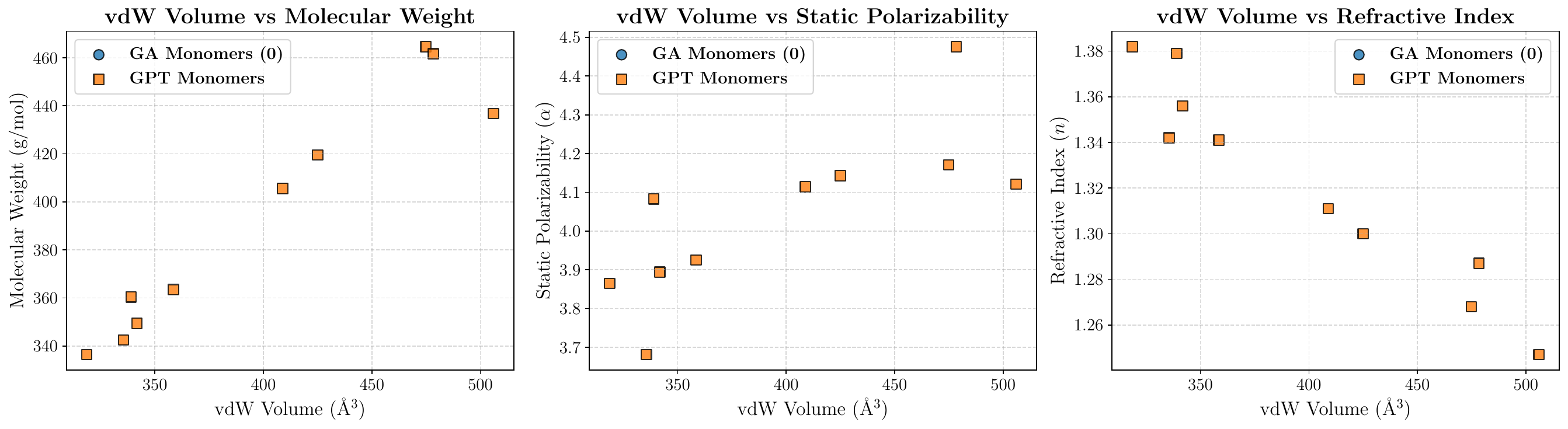}
    \caption{Comparison of vdW volume and (a) molecular weight, (b) static polarizability, and (c) refractive index for GA- and GPT-based models in Area D.}
    \label{fig:area_D}
\end{figure}

The training dataset in \emph{Area D} is populated with longer molecules that possess higher refractive indices. This characteristic hindered the GA model's ability to generate new molecules in this region. However, the GPT-based model successfully generates long molecules featuring cyanobiphenyl mesogens, ultra-long aliphatics, and extended multi-ring aromatics. It achieves this by combining flexible aliphatic spacers with rigid, highly conjugated cores to maximize volume and polarizability simultaneously. As shown in Fig.~\ref{fig:area_D}, the GPT-based model generates molecules with high vdW volumes ($300$ to $520~\AA^3$), high MW ($330$ to $470~\mathrm{g/mol}$), comparatively higher static polarizabilities ($3.6$ to $4.5$), and higher refractive indices ($1.24$ to $1.39$). This result also indicate that this area could be further explored aiming at the design of acrylates with high refractive index. 

\begin{figure}[!h]
    \centering
    \includegraphics[width=\linewidth]{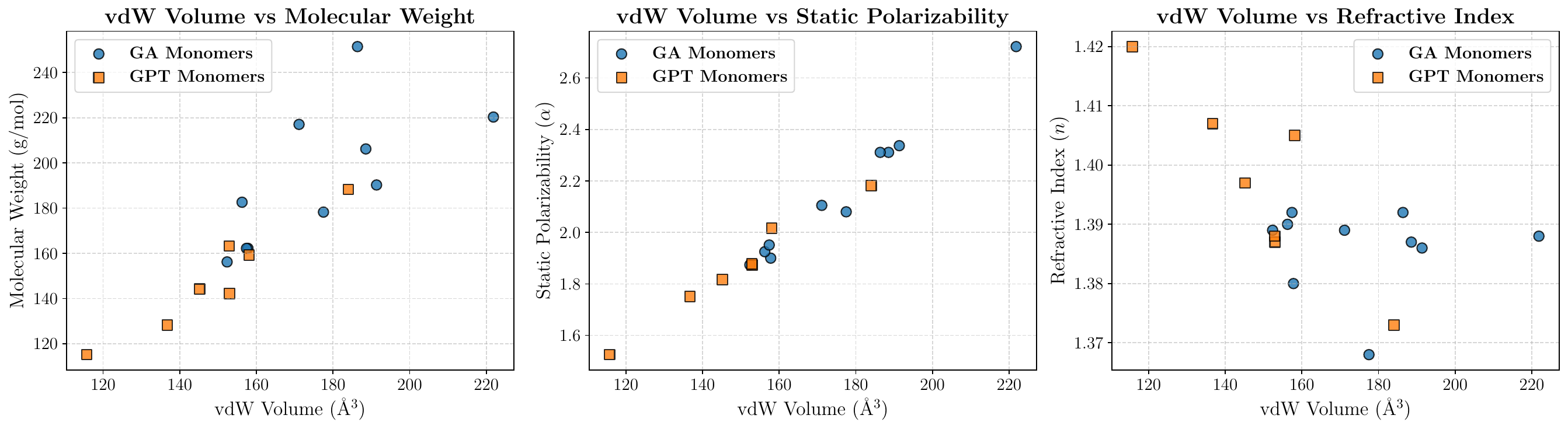}
    \caption{Comparison of vdW volume and (a) molecular weight, (b) static polarizability, and (c) refractive index for GA- and GPT-based models in Area F.}
    \label{fig:area_F}
\end{figure}

Neither the GA nor the GPT model was able to generate molecules in \emph{Area E}, where the training dataset is populated with dense aromatics featuring high refractive indices. Therefore, the incorporation of extra constraints on the chemical structure of the polymer of interest could remedy this deficiency. However, in \emph{Area F}, the GA model focuses on incorporating chlorinated aromatics and dense bridged polycycles to match the structural motifs of the training dataset, whereas the GPT model focuses on producing small aliphatics with polar heteroatoms (such as amines and hydroxyls) to mathematically fit the region's property constraints. As shown in Fig.~\ref{fig:area_F}, both models generate molecules across a wider range of vdW volumes ($110$ to $230~\AA^3$), molecular weights ($110$ to $250~\mathrm{g/mol}$), static polarizabilities ($1.5$ to $2.8$), and higher refractive indices ($1.36$ to $1.42$). Notably, within this area, the GA model tends to generate molecules with higher vdW volumes.

In summary, the analysis above reveals that while both models can target the same physical property regions, they employ fundamentally different structural strategies in the molecule generation. The GA model acts as a local exploiter, adhering closely to known training data motifs by permuting existing structures (e.g., modifying known fluorinated chains or dense bridged polycycles), which enables a more efficient filling of internal gaps in the latent space. In contrast, the GPT model acts as a broad structural explorer, demonstrating a unique ability to discover entirely novel chemical families---such as cyanobiphenyl mesogens, branched aliphatics, and linear hydrocarbon chains---that mathematically satisfy regional property constraints. This facilitates the generation of potentially long molecules possessing extreme physical properties (e.g., high refractive index).

\subsection{Comparison with state-of-the-art frameworks for polymer informatics}
To further contextualize the capabilities of \texttt{PolyGraphPy}, it is important to compare its performance against recent prominent frameworks addressing polymer inverse design under data constraints, such as the methodology proposed by Li et al. \cite{Li2026}. While both frameworks tackle the core challenge of discovering targeted polymers with limited data for specific applications, they present distinct operational profiles regarding generation efficiency, validity rates, and quantitative novelty. 

In terms of generation efficiency, Li et al. rely on Particle Swarm Optimization (PSO) to optimize feature vectors, followed by a depth-first search graph construction, ultimately validating the polymers using computationally expensive all-atom MD cooling simulations. In contrast, \texttt{PolyGraphPy} leverages a Bayesian GNN surrogate model that evaluates natively generated GPT and GA candidates in seconds, shifting the physical validation overhead from slow classical MD to rapid, automated DFTB+ quantum simulations. 

Regarding validity, Li et al. evaluate chemical viability by defining ``polymerizable molecules'' broadly (e.g., the presence of any double bond or bifunctional group). The proposed approach, however, explicitly enforces a highly restrictive and domain-specific constraint: the mandatory \texttt{C=C-C(=O)O} acrylate backbone. While this strict structural sieve yields a 12.6\% success rate for the GPT model, it provides a massive probabilistic enrichment over random sampling (0\%) and ensures absolute domain relevance. Finally, quantitative novelty metrics highlight the strong exploratory power of the dual-engine approach. Where recent studies like Li et al. evaluate novelty continuously through average Tanimoto similarity coefficients relative to the training data, \texttt{PolyGraphPy} achieves absolute novelty rates: 89.99\% (GA) and 78.23\% (GPT) of the generated candidates that are completely absent from the original training distribution. This direct comparison underscores that \texttt{PolyGraphPy} provides a highly competitive methodology, prioritizing rapid high-throughput screening and rigorous domain-specific absolute novelty for acrylate polymer discovery.

Moreover, \cite{Xu2026} introduced polyGT, a decoder-only GPT designed to generate new polymers via a self-supervised training process by learning the rules that lead to polymer structures with targeted properties. Notably, polyGT relies on text-based input via the Polymer Simplified Molecular Input Line Entry System (PSMILES), whereas \texttt{PolyGraphPy} employs a heterogeneous representation that combines SMILES, SELFIES, and graphs to provide a robust and detailed description of the polymer structure. Furthermore, polyGT's training dataset is constructed from approximately 18,000 PSMILES representations of synthetic homopolymers collected from the PolyInfo database, which is not publicly available. However, similar to \texttt{PolyGraphPy}, it incorporates RDKit to calculate several molecular properties (e.g., NH0, MolLogP, ring counts, aromatic rings, and NumHAcceptors) to label the data. Additionally, all-atom classical molecular dynamics (MD) simulations were used to compute the thermal conductivity ($T_c$) labels for 1,000 amorphous polymers. This represents a significant limitation compared to the approach employed in \texttt{PolyGraphPy}, which relies on highly efficient quantum mechanics calculations via DFTB.

Recently, the Text-to-Text Transfer Transformer (T5) architecture \cite{raffael2019} has been used to develop POLYT5 \cite{Sahu2026}, a domain-adapted encoder-decoder model for designing polymers (especially homopolymers) trained on 100 million chemically diverse polymer structures using the SELFIES representation. The objective of this framework is similar to that of \texttt{PolyGraphPy} in that it also focuses on both property prediction and molecule generation. However, POLYT5 was originally applied to dielectric polymer design, whereas \texttt{PolyGraphPy} focuses on optical properties. An important distinction between the two methods lies in the featurization process: while POLYT5 is primarily text-based, \texttt{PolyGraphPy} relies on graphs. This graph-based approach enables the rapid incorporation of copolymers, although the authors of POLYT5 note that their model could be extended to copolymers via fine-tuning on curated datasets, although this approach could not ensure an effective learning of the structure of the underlying manifold of the property space. Notably, during the prediction stage, POLYT5 exhibited large variability in its RMSE across the considered parameters, reporting RMSEs of 40.82 K, 67.07 K, and 78.59 K for glass transition temperature, melting temperature, and decomposition temperature, respectively. While a direct performance comparison with the predictive module of \texttt{PolyGraphPy} is not feasible here, the proposed non-Euclidean graph featurization in \texttt{PolyGraphPy} is expected to better capture complex data structures to achieve higher predictive accuracy. Additionally, whereas POLYT5 is strictly LLM-based, \texttt{PolyGraphPy} employs a heterogeneous architecture including both LLM and GA models. The advantage of \texttt{PolyGraphPy} in this regard lies in its ability to exploit both generation strategies, enabling the persistent filling of the latent space alongside extensive boundary exploration to generate polymers with extreme properties.

\section{Conclusions}

\texttt{PolyGraphPy}, an open-source, modular Python framework designed to streamline the discovery and design of polymeric materials, is introduced with a specific focus on acrylate monomers exhibiting tailored static polarizability. To address the scarcity of public datasets containing both molecular structures and static polarizability values, we generated two extensive datasets using DFTB+ simulations: (A) 3,427 monomer/homopolymer simulations and (B) 8,627 copolymer simulations. The construction of these large-scale datasets was made feasible by the computational efficiency of Density Functional Tight Binding (DFTB) compared to more expensive quantum mechanics methods, such as Density Functional Theory (DFT) or traditional Force Fields (FF). Specifically, by leveraging DFTB’s ability to approximate electronic structures using precomputed Slater-Koster (SK) parameters, datasets (A) and (B) were constructed in approximately 12 and 38 hours, respectively.

By integrating atomistic simulations with advanced predictive and generative models, \texttt{PolyGraphPy} addresses critical challenges in polymer informatics. The framework utilizes the computational power of DFTB+ to enable large-scale screenings and to train Bayesian Graph Neural Networks (GNNs). These GNNs, which utilize stochastic graph representations of polymers, are designed to predict material properties---such as static polarizability in the optical domain---while providing robust uncertainty quantification. We developed and trained two models based on the Graph U-Net architecture to predict the static polarizability of monomers/homopolymers and copolymers. Both models demonstrated high predictive accuracy: the monomer/homopolymer model achieved $\text{MAPE} = 11.83\%$, $R^2 = 0.9739$, and $\text{MSE} = 0.0015$, while the copolymer model achieved $\text{MAPE} = 5.19\%$, $R^2 = 0.9745$, and $\text{MSE} = 0.00093$.

Furthermore, \texttt{PolyGraphPy} enables efficient property-guided \textit{de novo} design via two complementary generative approaches: a SELFIES-based Generative Pretrained Transformer (GPT) and a Genetic Algorithm (GA) utilizing BRICS fragmentation. These models facilitate the exploration of the polymer chemical space to discover materials with targeted properties. The GA-based model generated 730 unique polymers, 89.99\% of which were entirely absent from the original acrylate dataset, highlighting its strong exploratory capacity. Notably, the physical validity of these generated structures was independently verified: DFTB+ simulations of the top 30 novel monomers generated by the GA demonstrated exceptional agreement with the GNN surrogate predictions, achieving an $R^2$ of 0.95 and a mean absolute error of 0.52 \AA{}$^3$. This rigorous validation confirms that the framework reliably discovers authentic, high-performing chemistries rather than exploiting adversarial artifacts within the predictive model. Similarly, the GPT-based model generated 126 valid monomers, of which 99.2\% were unique (125 distinct SMILES strings) and 78.23\% were novel compared to the training set, further demonstrating the model's ability to explore new chemical domains. As critically analyzed, while the raw generative yield of the GPT engine sits at roughly 12.6\%, its integration with automated structural filtering and high-speed Bayesian proxies ensures that novel, valid candidates are isolated in seconds, providing a highly practical tool for traversing constrained chemical spaces.

Due to its object-oriented and open-source philosophy, \texttt{PolyGraphPy} possesses a high level of customization and accessibility. It offers a unified, end-to-end pipeline that significantly reduces the time and computational costs associated with materials discovery, facilitating the systematic design of polymers with tailored architectures and functionalities.

\section*{Conclusions}
\texttt{PolyGraphPy}, an open-source, modular Python framework designed to streamline the discovery and design of polymeric materials, is introduced with a specific focus on acrylate monomers exhibiting tailored static polarizability. To address the scarcity of public datasets containing both molecular structures and static polarizability values, we generated two extensive datasets using DFTB+ simulations: (A) 3,427 monomer/homopolymer simulations and (B) 8,627 copolymer simulations. The construction of these large-scale datasets was made feasible by the computational efficiency of Density Functional Tight Binding (DFTB) compared to more expensive quantum mechanics methods, such as Density Functional Theory (DFT) or traditional Force Fields (FF). Specifically, by leveraging DFTB’s ability to approximate electronic structures using precomputed Slater-Koster (SK) parameters, datasets (A) and (B) were constructed in approximately 12 and 38 hours, respectively.

By integrating atomistic simulations with advanced predictive and generative models, \texttt{PolyGraphPy} addresses critical challenges in polymer informatics. The framework utilizes the computational power of DFTB+ to enable large-scale screenings and to train Bayesian Graph Neural Networks (GNNs). These GNNs, which utilize stochastic graph representations of polymers, are designed to predict material properties---such as static polarizability in the optical domain---while providing robust uncertainty quantification. We developed and trained two models based on the Graph U-Net architecture to predict the static polarizability of monomers/homopolymers and copolymers. Both models demonstrated high predictive accuracy: the monomer/homopolymer model achieved $\text{MAPE} = 11.83\%$, $R^2 = 0.9739$, and $\text{MSE} = 0.0015$, while the copolymer model achieved $\text{MAPE} = 5.19\%$, $R^2 = 0.9745$, and $\text{MSE} = 0.00093$.

Furthermore, \texttt{PolyGraphPy} enables efficient property-guided \textit{de novo} design via two complementary generative approaches: a SELFIES-based Generative Pretrained Transformer (GPT) and a Genetic Algorithm (GA) utilizing BRICS fragmentation. These models facilitate the exploration of the polymer chemical space to discover materials with targeted properties. The GA-based model generated 730 unique polymers, 89.99\% of which were entirely absent from the original acrylate dataset, highlighting its strong exploratory capacity. Similarly, the GPT-based model generated 126 valid monomers, of which 99.2\% were unique (125 distinct SMILES strings) and 78.23\% were novel compared to the training set, further demonstrating the model's ability to explore new chemical domains.

Due to its object-oriented and open-source philosophy, \texttt{PolyGraphPy} possesses a high level of customization and accessibility. It offers a unified, end-to-end pipeline that significantly reduces the time and computational costs associated with materials discovery, facilitating the systematic design of polymers with tailored architectures and functionalities.

\section*{CRediT authorship contribution statement}

\textbf{J. G. C. S. Duarte}: Writing – review \& editing, Writing – original draft, Validation, Methodology, Investigation, Formal analysis, Data curation,
Conceptualization. \textbf{S. Venkatram}: Validation, Methodology, Investigation, Formal analysis, Data curation. \textbf{M. Cencer}: Validation, Methodology, Investigation, Formal analysis, Data curation. \textbf{T. Dumitricǎ}: Writing – review
\& editing, Supervision, Project administration, Funding acquisition. \textbf{K. R. M. dos Santos}: Writing – review
\& editing, Supervision, Project administration, Funding acquisition.

\section*{Declaration of competing interest}
The authors declare the following financial interests/personal relationships, which may be considered potential competing interests:
K. R. M. dos Santos, T. Dumitricǎ, and J. G. C. S. Duarte report that financial support was provided by 3M. S. Venkatram and M. Cencer report employment by 3M. If there are other authors, they declare that they have no known competing financial interests or personal relationships that could have appeared to influence the work reported in this paper.

\section*{Acknowledgements}
The authors gratefully acknowledge the technical and financial assistance received from 3M (Award No. CON000000109024). Furthermore, the computational resources from the Minnesota Supercomputing Institute are greatly acknowledged.

\section*{Appendix A. Supplementary data}
Supplementary material related to this article can be found online.

\section*{Data availability}
The data that support the findings of this study are openly available in GitHub at \url{https://github.com/SASP-lab/PolyGraphPy}.

\bibliographystyle{model1-num-names}
\bibliography{mybibfile}

\end{document}